\documentclass{IEEEoj}
\usepackage{cite}
\usepackage{amsmath,amssymb,amsfonts}
\usepackage{algorithm}
\usepackage{algpseudocode}
\usepackage{graphicx,color}
\usepackage{textcomp}
\usepackage{soul}
\usepackage[english]{babel}
\usepackage{amsthm}
\usepackage{soul}

\newtheorem{theorem}{Theorem}
\newtheorem{lemma}{Lemma}
\def\BibTeX{{\rm B\kern-.05em{\sc i\kern-.025em b}\kern-.08em
    T\kern-.1667em\lower.7ex\hbox{E}\kern-.125emX}}
\AtBeginDocument{\definecolor{ojcolor}{cmyk}{0.93,0.59,0.15,0.02}}
\def\OJlogo{\vspace{-4pt}\hskip-4pt\includegraphics[height=18pt]{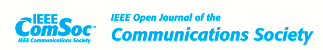}}
\usepackage{acronym}
\acrodef{rsm}[RSM]{receive spatial modulation}
\acrodef{adc}[ADC]{analog-to-digital converters}
\acrodef{ris}[RIS]{reconfigurable intelligent surfaces}
\acrodef{zf}[ZF]{zero-forzing}
\acrodef{ras}[RAS]{receiver antenna selection}
\acrodef{ara}[ARA]{active receiver antenna}
\acrodef{aras}[ARAs]{active receiver antennas}
\acrodef{dl}[DL]{downlink}
\acrodef{snr}[SNR]{signal-to-noise ratio}

\DeclareMathOperator{\Tr}{Tr}
\DeclareMathOperator*{\argmin}{arg\,min}

\DeclareUnicodeCharacter{2061}{}
\newcommand{\Hbu}{\mathbf{H}_{bu}}
\newcommand{\THbu}{\Tilde{\mathbf{H}}_{bu}}
\newcommand{\Hbr}{\mathbf{H}_{br}}
\newcommand{\THbr}{\Tilde{\mathbf{H}}_{br}}
\newcommand{\Hru}{\mathbf{H}_{ru}}
\newcommand{\THru}{\Tilde{\mathbf{H}}_{ru}}
\newcommand{\Ht}{\mathbf{H}}
\newcommand{\PhiM}{\mathbf{\Phi}}

\newcommand{\phiM}{\pmb{\varphi}}

\begin{document}
\receiveddate{13 February 2026}
\accepteddate{25 February 2026}
\publisheddate{2 March 2026}
\currentdate{24 March 2026}
\doiinfo{OJCOMS.2026.3669559}

\title{RIS-assisted Multiuser MISO Transmission and the Impact of Imperfect Channel Estimation}

\author{Ainna Yue Moreno-Locubiche\IEEEauthorrefmark{1}\IEEEmembership{(Student Member, IEEE)}, Josep Vidal\IEEEauthorrefmark{1}\IEEEmembership{(Senior Member, IEEE)}, Antonio Pascual-Iserte \IEEEauthorrefmark{1}\IEEEmembership{(Senior Member, IEEE)}, Olga Muñoz \IEEEauthorrefmark{1}\IEEEmembership{(Senior Member, IEEE)}}

\affil{Dept. of Signal Theory and Communications, Universitat Politècnica de Catalunya - BarcelonaTech (UPC), Spain}
\corresp{CORRESPONDING AUTHOR: Josep Vidal (e-mail: josep.vidal@upc.edu).}
\authornote{This work is part of the I+D+i project 6-SENSES (PID2022-138648OB-I00) funded by MICIU/AEI/10.13039/501100011033 and ERDF/EU and the grant 22CO1/008248.}
\markboth{Preparation of Papers for IEEE OPEN JOURNALS}{Author \textit{et al.}}

\begin{abstract}
This paper proposes the joint design of \ac{ris} and zero-forcing (ZF) precoding for the downlink (DL) multiuser multiple-input single-output (MU-MISO) setup in millimeter-wave (mmWave) bands, where ZF is particularly attractive due to its ability to suppress inter-user interference by exploiting the large antenna arrays and sparse directional channels characteristic of mmWave systems. This ensures efficient spatial multiplexing with manageable complexity, making ZF a practical and in modern 5G/6G deployments. However, a careful design is necessary to overcome potential rank deficiency in the channel matrix. For the MU-MISO case, rank deficiency may arise if users exhibit significantly different channel gains or if, being in far-field, they are aligned with the position of the transmitter. On the other hand, the deployment of a \ac{ris} introduces artificial scattering which can shape the radio environment to address those situations. We explore the joint design under perfect channel knowledge, assess the impact of imperfect channel estimation on the bit error rate (BER) and propose a robust design of pilot transmissions that equalizes multiuser interference across users in the presence of channel errors in the precoder design. This evaluation shows the advantages of optimized \ac{ris}-aided ZF MU-MISO communication for the DL of wireless systems.
\end{abstract}


\maketitle

\section{INTRODUCTION} \label{Intro}
The escalating demands for advanced wireless communication services are steering research towards the development of technologies that will ultimately shape the sixth-generation (6G) mobile communication standard. Key performance indicators such as spectral efficiency (SE), reliability, coverage, latency, jitter, and energy consumption are pivotal in this exploration. Among the considered technical solutions, the utilization of millimeter-wave (mmWave) bands stands out as the primary contender for achieving elevated transmission rates, thanks to the increased available bandwidth. However, the adoption of mmWave bands comes with challenges. These bands exhibit substantial propagation losses and limited indoor penetration when compared to centimeter wave bands. Moreover, reflections suffer high losses, leading to a significant reduction in multipath effects. Addressing these aspects is crucial for realizing the full potential of mmWave technology for 6G mobile standards.

Beyond mmWave transmission, multiple-input multiple-output (MIMO) also plays a crucial role in enhancing data rates \cite{Bjorn_mMIMO17, Larsson_2014}. Nevertheless, MIMO encounters two significant practical challenges. Foremost, it fails in achieving multiplexing gains in far-field line-of-sight (LOS) propagation environments due to the rank deficiency of MIMO channel matrices. Additionally, each receive active antenna element requires a radio frequency chain with its analog-to-digital converter, leading to a notably high total cost and energy consumption at the user equipment (UE) in the downlink (DL). This is particularly pronounced for massive MIMO systems, large bandwidths and a high number of quantization bits at the receivers \cite{WangMIMO16, XumMIMO18, XuMUmMIMO17}. Consequently, multiuser MISO setups \cite{6832894} are being strongly considered in DL multi-antenna transmissions for mmWave bands.

Zero Forcing (ZF) precoding is a widely proposed technique in MU-MISO DL transmission. This strategy efficiently enables complete spatial multiplexing and leverages multiuser diversity in high signal-to-noise ratio (SNR) environments \cite{1002659,1468466}, achieved by nulling multiuser interference (MUI) under the assumption of perfect channel state information (CSI). Nevertheless, rank deficiency of the MU-MISO channel matrix turns ZF unfeasible, a situation that may arise when:
\begin{itemize}
    \item channel attenuations of UEs engaged in MU-MISO transmission are very different because of distance or the presence of obstacles, or
     \item the two or more rows of the MU-MISO channel matrix are nearly proportional, a situation that appears when several UEs are aligned from the base station (BS) perspective and in the far-field of the BS antenna array.
\end{itemize}
These situations may be addressed through several mechanisms that include using minimum mean square error (MMSE) precoders (at the expenses of creating inter-user interference) or smartly grouping users (an NP-hard problem for which practical solutions have been proposed \cite{UserGroupingRubioIserte}). Beyond these approaches, we focus on exploring whether these situations are mitigated with a judicious design of reconfigurable intelligent surfaces (RIS).

Recently, \ac{ris}s have garnered considerable attention as a cost-effective and energy-efficient solution for intelligently shaping the radio environment by creating multipath scenarios. A smart design of \ac{ris} reflection coefficients can potentially improve penetration losses in buildings and shaded regions while simultaneously enhancing the rank of the single-user MIMO channel \cite{xu2022reconfiguring}. In scenarios where perfect CSI is available, the adjustment of \ac{ris} reflection coefficients at each reflecting element can be accomplished through the assistance of a smart controller. The capabilities of such a system have been extensively investigated across diverse settings \cite{wu2018intelligent, Zhang2020IRS, wu2018beamforming, Han2018LargeIS}, showcasing its efficiency in improving system throughput and coverage.

Beyond these foundational benefits, the role of \ac{ris} becomes even more prominent when viewed through the lens of emerging 6G use cases. The programmability and low‑power operation of \ac{ris} make them particularly attractive for highly dynamic environments such as UAV‑assisted communication, where controllable reflections can enhance air‑to‑ground links, mitigate blockages, and support energy‑efficient flight paths \cite{10420488}. In vehicular and V2X networks, \ac{ris} can reshape propagation around obstacles to ensure reliable, low‑latency connectivity for autonomous driving and cooperative perception \cite{10903998}. Similarly, in industrial IoT and smart factory deployments, \ac{ris} can provide robust coverage in harsh metallic environments while enabling high‑precision sensing and localization \cite{Li_2023}. These application scenarios illustrate how \ac{ris} extend beyond traditional coverage enhancement, positioning them as a key enabler for the programmable, environment‑aware communication paradigm envisioned for 6G networks.

Leveraging on those studies, we consider a multiuser communication system where a multi-antenna BS receives support from a RIS to transmit to a number of single antenna users. We analyse how the parameters of the RIS are optimized and what is the impact of imperfect channel estimation.

\subsection{RELATED WORKS}
This section reviews prior studies that frame our contribution, highlighting key advances and identifying gaps our work addresses. Recent research on RIS‑assisted multiuser systems has concentrated on two main fronts: channel estimation/feedback and precoding/beamforming optimization.

\textit{Channel Estimation}. A central challenge in RIS‑assisted MU‑MISO systems is the acquisition of accurate cascaded channel state information (CSI). Wang et al. \cite{wang_channel_2020} proposed frameworks and algorithms for multiuser channel estimation, analyzing pilot overhead and identifiability issues. Chen et al. \cite{9521836} extended this line of work by introducing hybrid evolutionary algorithms for sparse channel estimation in mmWave RIS‑assisted MIMO, exploiting sparsity to reduce training overhead and complexity. Nadeem et al. \cite{Nadeem20} provided one of the earliest integrated approaches, jointly considering channel estimation and beamforming design for MU‑MISO with RIS assistance. Shin et al. \cite{9771428} addressed the feedback bottleneck by proposing limited feedback schemes tailored to RIS‑aided MU‑MIMO. More recently, Lawal et al. \cite{11020694} investigated channel estimation for double‑RIS‑assisted multi‑user MIMO systems in obstructed environments, leveraging deep learning to optimize a two‑stage cascaded estimation framework and significantly improve accuracy under low‑SNR conditions. While these works advance estimation and feedback strategies, they evaluate estimation accuracy in isolation. None of them quantify how estimation errors propagate through fixed linear precoders, nor do they provide diagnostic insights into residual interference and rate degradation in MU‑MISO DL. Our work directly addresses this gap by modeling estimation errors and analyzing their impact on ZF robustness.

\textit{Precoding and Beamforming}. On the precoding side, Liu et al. \cite{Liu2019JointSP} investigated joint symbol‑level precoding and reflecting designs for RIS‑enhanced MU‑MISO, while Ur Rehman et al. \cite{9474428} proposed a vector approximate message passing‑based approach for joint active and passive beamforming in RIS‑assisted MU‑MIMO. Yu et al. \cite{YuZF22} studied RIS‑aided ZF and regularized ZF beamforming for simultaneous information and energy delivery, and Peel et al. \cite{1391204} provided classical baselines with vector‑perturbation and regularized channel inversion techniques for near‑capacity multiuser MIMO. More recently, Zhang et al. \cite{10508095} introduced a multi‑agent reinforcement learning framework for joint precoding and RIS phase‑shift optimization in cell‑free massive MIMO systems, demonstrating how distributed learning agents can effectively coordinate active and passive beamforming to enhance scalability and performance in large‑scale deployments. These contributions demonstrate the potential of joint optimization and advanced precoding strategies under ideal CSI assumptions. However, they do not isolate the performance of ZF precoding under imperfect CSI. In contrast, our proposal focuses on the robustness of ZF in RIS‑assisted MU‑MISO, explicitly quantifying how estimation errors degrade BER and sum‑rate/capacity, evaluating how RIS mitigate this errors, and identifying scenarios where ZF remains viable. Additionally, we show that the adoption of ZF precoding allows uncoupling the precoder and the RIS designs, which cannot be claimed for MMSE precoding.

\textit{Imperfect CSI}. 
Several works have examined the impact of non‑idealities on RIS‑assisted systems. Pérez‑Adán et al. \cite{Pérez-Adan_AlternatingMinimization} studied wideband MU‑MIMO under imperfect CSI using alternating minimization, while Lu et al. \cite{Zhang23} analyzed RIS‑assisted communications with hardware impairments and channel aging. Zhang et al. \cite{Zhang_SumRateMax} considered sum‑rate maximization under statistical CSI assumptions. These studies highlight performance degradation due to imperfect CSI and hardware limitations, but they remain solver‑centric and do not explicitly connect error statistics to the robustness of fixed precoding schemes. They also focus on optimization under statistical or approximate CSI rather than diagnostic evaluation of error propagation. Our work complements these efforts.

Taken together, these works emphasize optimization and architectural innovation, but a clear gap persists. Throughout our study, we reinforce the need for robust RIS optimization techniques and validation of RIS-assisted transmission performance under realistic rank-deficient conditions.

\subsection{CONTRIBUTION}
The paper deals with the use of RIS to improve the received power in a MU setup in the DL. The contribution of our study is as follows:
\begin{itemize}
    \item We show that the ZF structure at the transmitter allows uncoupling the design of the RIS coefficients and the precoder and hence the use of alternating optimization, as it is customary in many other approaches like \cite{8855810}, is no longer needed. 
    \item In the lack of a closed form solution for the RIS coefficients, we define a gradient-based design of the RIS coefficients that maximizes the received SNR, using complex matrix differentiation theory. This approach was adopted in \cite{10437329} in the context of single-user MIMO receive spatial modulation (RSM) transceivers.
    \item As channel estimation errors generate  interference (MUI) and performance degradation, our secondary objective is to assess the impact of using RIS in scenarios where the channel is imperfectly estimated. We propose a robust method that equalizes the MUI across users.
    \item We determine by simulations the beneficial impact of passive RISs in challenging multiuser scenarios with rank deficient channel matrix, that include signal obstruction between BS and UE, alignment of UEs with the BS in far-field, and substantial differences in channel gain across BS-UE links engaged in a multiuser transmission.
\end{itemize}
Results reveal insights into the efficacy and robustness of passive RIS deployment in ZF MU-MISO DL communication for users in shadowed and non-shadowed areas, and when they are aligned with respect to the BS.

The paper is organized as follows. The signal model for a MU-MISO transmission using ZF-precoding is introduced in section \ref{MU-MISO signal model}. The channel model including RIS is described in section \ref{Channel model}. The gradient-based algorithm used in the optimization of the phases of the passive RIS is in section \ref{Gradient optimization}. The least-squares channel estimation error is presented in section \ref{sec:channel_estimation} along with an statistical evaluation of the estimation error and the tradeoffs in spectral efficiency associated with pilot transmission for channel estimation. The impact of the imperfect channel estimation is analysed in terms of the BER in section \ref{sec:channel_unc}. Simulations are provided in section \ref{Results} to support our claims. Conclusions are included in section \ref{conclusions}.
\subsection{NOTATION}
The following notation is used throughout the paper. Boldface lower-case and upper-case characters denote vectors and matrices respectively: $\mathbf{x}\in \mathbb{C}^{N\times 1}$ and $\mathbf{X}\in \mathbb{C}^{M\times N}$ represent a vector of size $N$ and a matrix of size $M\times N$ respectively with complex entries. The superscripts $(\cdot)^*$, $(\cdot)^T$ and $(\cdot)^H$ represent the conjugate, transpose and conjugate transpose respectively. $\mathbb{E\{\cdot\}}$ represents the expectation operator. $\mathrm{Tr}\{\mathbf{X}\}$ is the trace of matrix $\mathbf{X}$.  The operator $\text{vec}(\mathbf{X})$ rearranges the elements of  matrix $\mathbf{X}$ column-wise. $\left \| \mathbf{x} \right \|$ stands for the Euclidean norm of vector $\mathbf{x}$. The operator $\text{diag}(\mathbf{x})$ is the $N\times N$ diagonal matrix whose entries are the $N$ elements of the vector $\mathbf{x}$. $\operatorname{Im}(\mathbf{x})$ stands for the imaginary part of vector $\mathbf{x}$. $\hat{\mathbf{X}}$ is the estimator of the random matrix $\mathbf{X}$. $\mathbf{X} \otimes \mathbf{Y}$ is the Kronecker product between $\mathbf{X}$ and $\mathbf{Y}$. The entry of a matrix $\mathbf{X}$ in the $i$-th row and the $j$-th column is expressed as $\left[\mathbf{X}\right]_{i,j}$. A random vector $\mathbf{x}\sim \mathcal{CN}(\mathbf{m},\mathbf{C})$ is circularly symmetric complex Gaussian distributed with mean vector $\mathbf{m}$ and covariance matrix $\mathbf{C}$.

\section{CHANNEL MODEL INCLUDING \ac{ris}} \label{Channel model}
\begin{figure}
\begin{center}
\includegraphics[scale=0.45]
{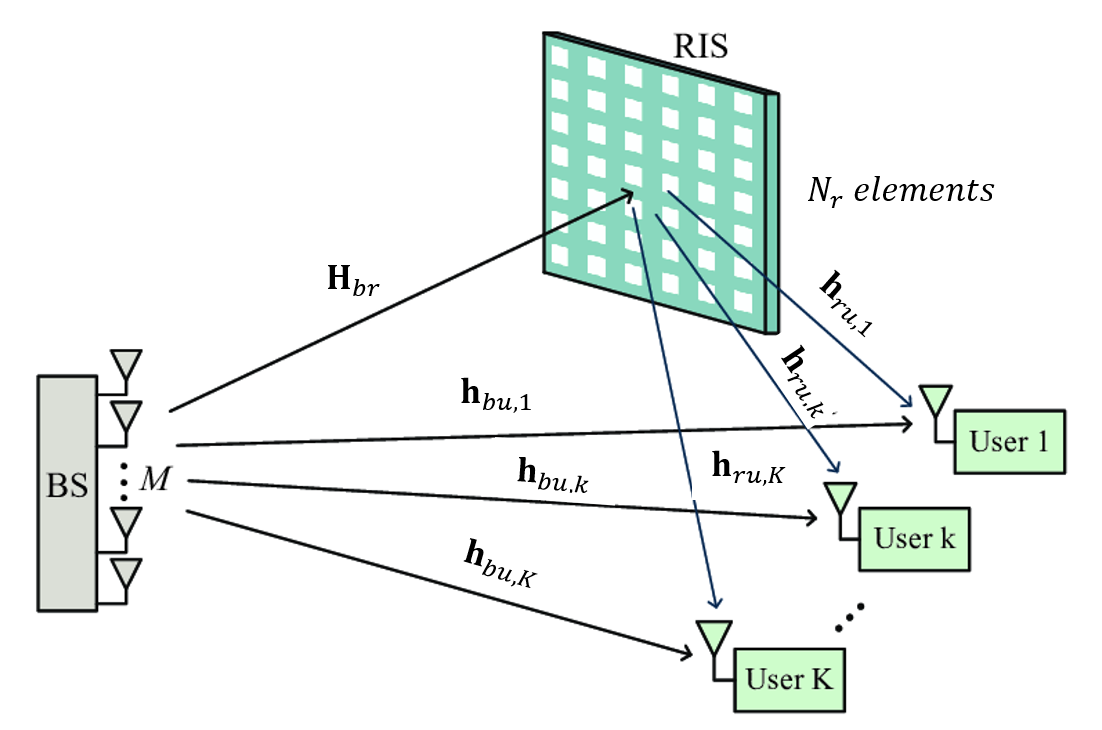}
    \caption{Channel model for the \ac{ris}-assisted DL transmission \cite{Liu2019JointSP}.}
    \label{fig:channel_model}
\end{center}
\end{figure}

We consider a MISO wireless system for multiuser DL transmissions, including the presence of an \ac{ris} in the scenario, whose $N_r$ elements are purely passive reflectors. The MU-MISO channel $\mathbf{H}$ shown in Fig. \ref{fig:channel_model} is expressed as:
\begin{equation}
   \label{eq:channel_model_RIS} 
   \mathbf{H}=\Hbu+\mathbf{H}_c \in \mathbb{C}^{K\times M}.   
\end{equation}
$\Hbu  \in \mathbb{C}^{K\times M}$ contains the direct channel gains between BS and the $K$ single-antenna users and $\mathbf{H}_c$ is the compound channel linking the BS, \ac{ris}, and users, which may be written as:
\begin{equation}
    \label{eq:compund_channel}
    \mathbf{H}_c=\Hru\PhiM\Hbr,
\end{equation}
where $\Hbr\in\mathbb{C}^{N_r\times M}$ contains the channel gains between the BS and the \ac{ris} and $\Hru\in\mathbb{C}^{K\times N_r}$ contains the channel gains between the \ac{ris} and the $K$ users. Thus, the $k$-th row of $\Hru$, $\mathbf{h}_{ru,k}^T\in\mathbb{C}^{1\times N_r}$, denotes the channel between the \ac{ris} and the $k$-th user. For a passive diagonal \ac{ris}, the unit-modulus reflection coefficients are stacked in $\PhiM=\text{diag}(\pmb{\varphi})=\text{diag}(e^{j\pmb{\phi}})\in \mathbb{C}^{N_r\times N_r}$, where $\pmb{\phi}=\left [\phi_1 \: \phi_2 \: \cdots \: \phi_{N_r}  \right ]$ contains de phases of the reflection coefficients of the $N_r$ RIS elements.

The channel coefficients for the BS-UE$_k$ link are:
\begin{equation}
    \label{eq:direct_channel_su}      [\mathbf{H}_{bu}]_{k,l}=\sqrt{\beta_{bu,k,l}}\exp{\left(-j \frac{2\pi f_c}{c}d_{k,l}\right)}, \:\:\:\:l=1,\dots,M,
\end{equation}
where $\beta_{bu,k,l}$ is a pathloss component to be discussed in Sec. \ref{Results}.A, $f_c$ is the carrier frequency, $d_{k,l}$ is the distance between the $l$-th antenna at the BS and the $k$-th UE, and $c$ is the speed of light. Likewise, the elements of $\mathbf{H}_{ru}$ and $\Hbr$ are:
\begin{equation}
    \label{eq:channel_ru}
        [\mathbf{H}_{ru}]_{k,i}=\sqrt{g_{ru,k,i}}\exp{\left(-j \frac{2\pi f_c}{c}d_{k,i}\right),}
\end{equation}
\begin{equation}
    \label{eq:channel_br}
        [\Hbr]_{i,l}=\sqrt{g_{br,i,l}}\exp{\left(-j \frac{2\pi f_c}{c}d_{i,l}\right)},
\end{equation}
where $g_{ru,k,i}$ and $g_{br,i,l}$ are pathloss components to be discussed in Sec. \ref{Results}.A, $d_{k,i}$ is the distance between the $i$-th element at the RIS and the $k$-th UE, and $d_{i,l}$ is the distance between the $i$-th element at the RIS and the $l$-th BS antenna.

\section{MU-MISO SIGNAL MODEL} \label{MU-MISO signal model}

The BS uses $M$ antennas to transmit data symbols $\mathbf{s} = \begin{bmatrix} s_1 & s_2 & \cdots & s_K
\end{bmatrix}^T$, that take independent zero-mean unit-power random complex values, to $K\leq M$ single-antenna users. The signal received by the $k$-th user at a time instant $t$ is given by:
\begin{equation}
    y_{t,k}=\alpha \mathbf{h}_{k}^T\mathbf{B}\mathbf{\Gamma}^\frac12\mathbf{s}_{t} + n_{t,k},
    \label{eq:signal_model}
\end{equation} 
where $\mathbf{h}_k^T\in\mathbb{C}^{1\times M}$ is the propagation channel from the $M$ BS antennas to the $k$-th user, $\mathbf{B}\in\mathbb{C}^{M\times K}$ is the precoding matrix, $\mathbf{\Gamma}=\text{diag}(\gamma_1,\cdots,\gamma_K),\: \gamma_i>0$ accounts for a pre-fixed balancing on the per-user transmitted power, and $\alpha$ is a normalization factor ensuring that the total transmitted power is limited. On the other hand, $n_{t,k}$ is the zero-mean stationary i.i.d. circularly symmetric complex Gaussian random noise with variance $\sigma_n^2$. The multiuser channel matrix $\Ht \in \mathbb{C}^{K \times M}$ is defined as $[\Ht]_{k,:} = \mathbf{h}_k^T$.
On the transmit side, channel knowledge assumption allows defining $\mathbf{B}$ as:
\begin{equation} 
    \mathbf{B}=\Ht^{H}(\Ht\Ht^{H})^{-1},
    \label{eq:zf}
\end{equation}
which is the optimum linear ZF precoding choice for maximum received signal and maximum fairness under total transmit power constraint \cite{4599181}. Then, the received signal at the $k$-th UE active antenna can be rewritten as:
\begin{equation}
    \label{eq:y_model}
    y_{t,k}=\alpha\gamma_k^\frac12 s_{t,k} + n_{t,k}.
\end{equation}
Given a transmitted vector signal $\mathbf{x}_t=\alpha\mathbf{B}\mathbf{\Gamma}^\frac12\mathbf{s}_{t}$, we can express the transmit power $P_T$ as:
\begin{equation} 
    P_T=\Tr\left \{\mathbb{E} \left \{\mathbf{x}_t\mathbf{x}_t^H\right\}\right \}=\alpha^2\mathrm{Tr}\left\{ \left ( \mathbf{H}\mathbf{H}^{H} \right )^{-1} \mathbf{\Gamma}\right\}.
    \label{eq:transmitted_power}
\end{equation}
Setting the total transmit power to $P_T=P$, we obtain:
\begin{equation} 
    \alpha^2=P\:\mathrm{Tr}\left\{\left( \mathbf{H}\mathbf{H}^{H} \right )^{-1} \mathbf{\Gamma}\right\}^{-1},
    \label{eq:norm_factor}
\end{equation}
and $P_{r,k}=\alpha^2\gamma_k$ is the received signal power for user $k$. Therefore, to maximize the received power on each user, we define the following optimization problem:

\begin{equation}
\label{eq:optim_problem}
\begin{aligned}
\max_{\pmb{\phi}} \quad & \mathrm{Tr}\!\left\{\left( \mathbf{H}\mathbf{H}^{H} \right )^{-1} \mathbf{\Gamma}\right\}^{-1} \\
\text{s.t.} \quad & \phi_i \in [0, 2\pi), \ \forall i = 1, \dots, N_r.
\end{aligned}
\end{equation}

The optimum value of $\alpha^2$ is limited by the least eigenvalue of $\mathbf{H}\mathbf{H}^H$ in this way (the proof is in Appendix \ref{Bound1 on alpha}):
\begin{theorem}
    The value of $\alpha^2$ is upper bounded as:\begin{equation}
\alpha^2\leq P\:\lambda_{min}\left(\mathbf{H}\mathbf{H}^{H}\right)/\gamma_{min}.
\end{equation}
\end{theorem}
Consequently, deficiency in the rank of $\mathbf{H}$ drastically reduces the maximum received power. Our ambition is to increase the minimum singular value of the MU-MISO channel thanks to the multipath induced by the optimized \ac{ris}, and select the reflection coefficients to maximally enhance the received power for all $K$ users. 

\section{JOINT DESIGN OF PRECODER AND \ac{ris} PHASES}\label{Gradient optimization}
In the lack of an analytic solution to the problem in \eqref{eq:optim_problem}, we propose to use a gradient-based approach. Since $\alpha$ is a periodic function with $\pmb{\phi}$ and has a continuous derivative, we cannot claim convexity and applying gradient optimization may return suboptimal solutions. In practice, convergence to a sensible solution is consistently observed for the setups described in the results in section VII.

We adopt $J=\Tr\{(\Ht\Ht^H)^{-1}\mathbf{\Gamma}\}$ as the objective function to be minimized by computing its differential with respect to $\pmb{\phi}$. We adopt the complex matrix differentiation framework in \cite{Hjrungnes2011ComplexValuedMD} to obtain the gradient (see details in Appendix \ref{gradient expression}):
\begin{equation}
\label{eq:gradient}
    \!D_{\pmb{\phi}}J=2\operatorname{Im}\{\text{vec}((\Ht\Ht^H)^{-1}\mathbf{\Gamma}(\Ht\Ht^H)^{-1}\Ht)^H \mathbf{H}_{12}\textup{diag}(\pmb{\varphi})\}.
\end{equation}
The matrix
\begin{equation}
\label{eq:H12}
 \mathbf{H}_{12}=\begin{bmatrix}
\text{vec}(\mathbf{h}_{ru,1}\mathbf{h}_{br,1}^{T}) & \cdots  & \text{vec}(\mathbf{h}_{ru,N_r}\mathbf{h}_{br,N_r}^{T})
\end{bmatrix}
\end{equation}
contains $\mathbf{h}_{ru,i}$ and $\mathbf{h}_{br,i}$, defined as the $i$-th column of the matrices $\mathbf{H}_{ru}$ and $\mathbf{H}_{br}$. In our simulations, we chose the non-linear conjugate gradient method for faster convergence \cite{shewchuk1994introduction}. Once converged, the ZF precoder is obtained from \eqref{eq:zf}. Algorithm 1 shows the full procedure. The bulk of the complexity is in the inversion of $\Ht\Ht^H\in\mathbb{C}^{K\times K}$ and the products $(\Ht\Ht^H)^{-1}\Ht$ and $\mathbf{H}_{12}\text{diag}(\pmb{\varphi})$ per iteration. The number of complex products is $\mathcal{O}(K^3)+\mathcal{O}(MK^2)+\mathcal{O}(KMN_r)$, and convergence typically requires  10 to 15 iterations. 

Compared with other methods like \cite{Liu2019JointSP}, our algorithm offers a substantially more scalable computational profile as it does not exhibit exponential dependence on $N_r$ or $M$. The approach in~\cite{Liu2019JointSP} not only involves a baseline complexity of $\mathcal{O}\!\left((2N_r)^{1.5}\right)$ per iteration of the alternate optimization, but also incurs an additional cost due to the joint optimization of the RIS and the precoder of $\mathcal{O}\!\left(\Omega^K M^{3}\right)$, where $\Omega$ is the number of constellation symbols and $\Omega^K$ the amount of precoder vectors to be optimized. The computational complexity of our approach is similar to the fast version of the MU-MISO optimization approach shown in \cite{ebrahimi2025}, which is $\mathcal{O}(MK^2)+\mathcal{O}(KMN_r)$.

\begin{algorithm}
\label{alg:2}
\caption{Optimization through gradient iteration}
\begin{algorithmic}
\State $\textbf{input: } \mathbf{H}_{bu}, \mathbf{H}_{br}, \mathbf{H}_{ru}, \mathbf{\Gamma}$ 
\State $\textbf{output: } \pmb{\varphi},\mathbf{B}$ 
\State $compute \:\mathbf{H}_{12}$ \:\:\:\:\:\:\:\:\:\:\:\:\:\:\:\:\:\:\:\:\:{\small \text{ \% Defined in \eqref{dvec(H)}}}
\State $t=0,\:J_0=0,\:\pmb{\phi}_0\gets random,\:\pmb{\varphi}_0=e^{j\pmb{\phi}_0}$
\Repeat
\State $\mathbf{H}=\Hbu+\Hru \textup{diag}(\pmb{\varphi}_t)\Hbr$
\State $t \gets t+1$
\State $J_t=\mathrm{Tr}\{ \left ( \mathbf{H}\mathbf{H}^{H} \right )^{-1}\mathbf{\Gamma}\}$
\State $\mathbf{g}_t= -D_{\pmb{\phi}}J_{t-1}\:\:\:\:\:\:\:\:\:\:\:\:\:\:\:\:\:\:\:\:\:${\small \text{\% Minus gradient in \eqref{eq:gradient}}}
\State $\beta=\max\left\{\mathbf{g}_t^T(\mathbf{g}_t-\mathbf{g}_{t-1})(\mathbf{g}_{t-1}^T\mathbf{g}_{t-1})^{-1},0\right\}$
\State $\mathbf{u}_t=\mathbf{g}_t+\beta\mathbf{u}_{t-1}\:\:\:\:\:\:\:\:\:\:\:\:\:\:\:\:${\small \text{\% Conjugate gradient vector}}
\State $\mu = \argmin_\mu J_t|_{\pmb{\phi}_t+\mu\mathbf{s}_t}$ \: {\small \text{\% Conjugate gradient step size}}
\State $\pmb{\phi}_t=\pmb{\phi}_{t-1}+\mu\mathbf{u}_t$\:\:\:\:\:\:\:\:\:\:\:\:\:\: {\small \text{\% Gradient descend update}}
\State $\pmb{\varphi}_t=e^{j\pmb{\phi}_t}$
\Until {$|\frac{J_t-J_{t-1}}{J_t}|<\epsilon$}
\State $\mathbf{B}=\Ht^{H}(\Ht\Ht^{H})^{-1}$
\end{algorithmic}
\end{algorithm}

\section{CHANNEL ESTIMATION ERRORS}
\label{sec:channel_estimation}
Our goal in this section is, rather than proposing a practical and original channel estimation procedure, to identify a method that 1) provides expressions of the variance of the channel estimates, and 2) is able to estimate the different channel components that are present in the gradient in equation \eqref{eq:gradient} using the conventional transmission of uplink pilots coordinated with phases at the RIS.

Having a closed-form expression of the variance of channel estimates will allow to analytically define how estimation errors propagate into precoder design and degrade system performance. Among the many publications on channel estimation for RIS transmissions (e.g., \cite{wang_channel_2020, zheng_efficient_2021, zheng_intelligent_2020, zheng_uplink_2021}), we adopt the least squares (LS) approach \cite{Nadeem20}\cite{9130088}, which provides unbiased estimates and allows an analytic expression of the variance (the derivation is in Appendix \ref{Channel estimation}).

When the channel exhibits structure due to LOS components or other statistical knowledge is available \cite{9864300}, MMSE, Bayesian or parametric estimators can achieve lower variance and thereby outperform LS, as long as the model fits reality.

Although LS estimation offers a simple baseline for RIS-assisted systems, it assumes channel coefficients are independent random variables and therefore requires long training and is sensitive to noise, making it impractical for large-scale deployments. In this respect, the variances obtained from the adoption of LS may be considered as upper bounds. We do not adopt MMSE estimation here, since the presence of the regularization term prevents from obtaining closed-form expressions. More practical approaches such as ON/OFF protocols are hardware-wise friendly yet inefficient due to non-orthogonality and high pilot overhead \cite{wang_channel_2020}, while compressed sensing reduces training length by exploiting sparsity at the cost of added computational complexity \cite{10053657}. Finally, discrete RIS configurations stem from hardware limitations, with each element restricted to a finite set of phase shifts or reflection states. These constraints reduce design flexibility and demand tailored orthogonal patterns to balance feasibility and performance \cite{Nadeem20,11139112}.

We therefore assume that the channel components in \eqref{eq:channel_model_RIS} are estimated at each antenna element of the BS with the help of orthogonal pilots symbols transmitted in the UL by each UE. If channel reciprocity holds, the channel estimated in the UL $\Tilde{\mathbf{H}}$ is the transpose of the channel in DL,  $\Tilde{\mathbf{H}}= \mathbf{H}^T$. 

We define $\mathbf{V}_k=\left[\Tilde{\mathbf{h}}_{bu,k} \: \:\:\Tilde{\mathbf{H}}_{br}{\text{diag}}(\Tilde{\mathbf{h}}_{ru,k})\right] \in \mathbb{C}^{M \times (N_r+1)}$ as the matrix containing all the propagation channels for user $k$ involved in the optimization of the RIS: $\mathbf{V}_k$ encapsulates the elements required both in the gradient $D_{\pmb{\phi}}$ and $J$. 

As detailed in Appendix \ref{Channel estimation}, once $\mathbf{V}_1, \dots, \mathbf{V}_K$ have been estimated from symbols received at the BS, the full channel matrix $\mathbf{H}$, that includes the RIS elements, can be reconstructed. Moreover, the matrix $\mathbf{H}_{12}$ appearing in \eqref{eq:H12} can be derived by rearranging the elements of the estimated $\mathbf{V}_1, \dots, \mathbf{V}_K$. Arranging the elements of the estimated $\hat{\mathbf{V}}_k$ into a vector and using the expression of the received signal in \eqref{eq:ChannelModelUL}, \eqref{eq: rxSignalUplink}, and \eqref{least_squares_estimation}, we obtain:
\begin{equation}
\begin{aligned}
    &\text{vec}(\hat{\mathbf{V}}_k)=\frac{1}{\sqrt{P_{UL}}}\left(\left(\mathbf{\Omega}^*\mathbf{\Omega}^T\right)^{-1}\mathbf{\Omega}^*\otimes\mathbf{I}_M\right)\text{vec}\left(\mathbf{Y}_k\right)\\&=\text{vec}(\mathbf{V}_k)+\frac{1}{\sqrt{P_{UL}}}\left(\left(\mathbf{\Omega}^*\mathbf{\Omega}^T\right)^{-1}\mathbf{\Omega}^*\otimes\mathbf{I}_M\right)\text{vec}\left(\mathbf{N}_k\right),
\end{aligned}
\end{equation}
where $P_{UL}$ is the power transmitted in the UL, $\mathbf{\Omega}$ is the matrix containing pilots as defined in \eqref{eq: rxSignalUplink}, and $\mathbf{N}_k$ contains the noise observed in the UL. Assuming zero-mean circular white Gaussian noise of power $\sigma^2$ in $\mathbf{N}_k$, the estimate $\text{vec}(\hat{\mathbf{V}}_k)$ is unbiased and its covariance matrix is:
\begin{equation}
\label{eq:cov_mat} 
\begin{aligned}
\mathbf{C}_{\text{vec}(\mathbf{\hat{V}}_k)} &= \frac{\sigma^2}{P_{UL}T}\left(\mathbf{\Omega}^*\mathbf{\Omega}^T\right)^{-1}\otimes\mathbf{I}_M
\\&=\frac{\sigma^2}{P_{UL}T}\mathbf{I}_{M(N_r+1)},\:\: \text{with } T\geq N_r+1.
\end{aligned}
\end{equation}
If $\mathbf{\Omega}$ is designed such that its Grammian is the identity, the variance of the estimate is minimum for a given power of pilots. The Fourier and the Hadamard matrices meet this condition \cite{Nadeem20}.

As stated in Appendix \ref{Channel estimation}, the number of pilot symbols required to estimate the channel for user $k$ is $(N_r+1)$. Plugging $\hat{\mathbf{V}}_k$ into the channel model in \eqref{eq:H_Vk} we obtain:
\begin{multline}
\hspace{-0.3cm}\label{eq:H_estimated}
   \hat{\mathbf{H}}=\left (
    \begin{bmatrix}
        \mathbf{V}_1^{:,1} & \cdots & \mathbf{V}_K^{:,1}
    \end{bmatrix}^{T} + \mathbf{E}^{(1)}\right )\\ + (\mathbf{I}_M \otimes \phiM)^{T}
   \left (
    \begin{bmatrix}
        \mathbf{V}_1^{:,2:(N_r+1)} & \cdots & \mathbf{V}_K^{:,2:(N_r+1)}
    \end{bmatrix}^{T} \hspace{-0.1cm}+ \mathbf{E}^{(2)}\hspace{-0.1cm}\right ) ,
 \end{multline}
where $\mathbf{E}^{(1)}$ is the matrix containing the estimation error in the first columns of $\mathbf{V}_1,\dots,\mathbf{V}_K$, and $\mathbf{E}^{(2)}$ is the matrix containing the estimation error in the remaining $N_r$ columns. From this expression we can characterize the error in the estimated channel, which will be needed in Sec. \ref{sec:channel_unc}. Because of the uncorrelation obtained in \eqref{eq:cov_mat} among the elements of $\mathbf{V}_k$, all elements of $\mathbf{E}^{(1)}$ and of $\mathbf{E}^{(2)}$ are uncorrelated.

Equation \eqref{eq:cov_mat} shows that the variance in the estimated channels can be reduced if increasing the energy devoted to pilots by either increasing the transmission power in the UL or by using more pilot symbols $T$. This second option will entail a reduction of the spectral efficiency. 

\section{IMPACT OF CHANNEL ESTIMATION ERRORS ON BER}
\label{sec:channel_unc}
The BER analysis is adopted as a performance indicator in the evaluation of the system performance and the impact of channel estimation errors. By quantifying the BER under both perfect and imperfect estimation, we will gain insight into how estimation inaccuracies degrade signal detection for the RIS-based MU-MISO deployment.

 We assume that the channel uncertainty comes exclusively from the estimation errors in the UL, implying a difference between $\mathbf{\hat{H}}$ used in the ZF precoder $\mathbf{\hat{B}}=\mathbf{\hat{H}}^H(\mathbf{\hat{H}}\mathbf{\hat{H}}^H)^{-1}$ and the actual propagation channel $\Ht$. The received signal in the DL is, therefore:
\begin{equation}
    \label{eq:signal_model_error}
    y_{t,k}=\hat{\alpha} \mathbf{h}_{k}^T\mathbf{\hat{B}}\mathbf{\Gamma}^\frac12\mathbf{s}_{t} + n_{t,k},
\end{equation}
and the total transmitted signal power is:
\begin{equation}
    P_T=\hat{\alpha}^2\mathrm{Tr}\left\{\left( \mathbf{\hat{H}}\mathbf{\hat{H}}^{H}  \right)^{-1} \mathbf{\Gamma}\right\}.
\end{equation}
The received signal for user $k$ is disturbed by channel estimation errors present in $\hat{\mathbf{B}}$ and $\hat{\alpha}$ in this way:
\begin{equation}
\begin{split}
    \label{eq:signal_model_error_userk}
   y_{t,k}^\varepsilon &=(\alpha+d{\alpha}) \mathbf{h}_{k}^T(\mathbf{B}+d\mathbf{B})\mathbf{\Gamma}^\frac12\mathbf{s}_{t} + n_{t,k} \\
    &\backsimeq\hat{\alpha} \gamma_k^\frac12{s}_{t,k}+\alpha  \mathbf{h}_{k}^Td\mathbf{B}\mathbf{\Gamma}^\frac12\mathbf{s}_{t} + n_{t,k},
\end{split}
\end{equation}
where the term that includes the product of differentials $d{\alpha}\mathbf{h}_k^Td\mathbf{B}\mathbf{\Gamma}^\frac12\mathbf{s}_{t}$ has been dropped in the approximation. An expression for $d\mathbf{B}$ is obtained from complex calculus \cite{Hjrungnes2011ComplexValuedMD}:
\begin{equation*}
    \mathbf{\hat{B}} = \mathbf{\hat{H}}^H\left(\mathbf{\hat{H}}\mathbf{\hat{H}}^H\right)^{-1}=\mathbf{B}+d\mathbf{B},
\end{equation*}
\begin{equation*}
    d\mathbf{B} = -\mathbf{B}d\Ht\mathbf{B} + \mathbf{B}\mathbf{B}^Hd\Ht^H(\mathbf{I}-\Ht\mathbf{B})  + (\mathbf{I} - \mathbf{B}\Ht)d\Ht^H\mathbf{B}^H\mathbf{B},
\end{equation*}
\begin{equation}
    \label{eq:HdB}
    \Ht d\mathbf{B} = -d\Ht\mathbf{B} + \Ht(\mathbf{I} - \mathbf{B}\Ht)d\Ht^H\mathbf{B}^H\mathbf{B} = -d\Ht\mathbf{B},
\end{equation}
If the differential $d\Ht$ is a assumed to be the disturbance on $\mathbf{H}$ and identified as the channel estimation error $\mathbf{E}$, we can use \eqref{eq:signal_model_error_userk} and \eqref{eq:HdB}, to obtain an expression for the received signal at the $k$-th UE:
\begin{equation}
    \label{eq:y_epsilon}
\begin{aligned}
    \!&y_{t,k}^\varepsilon=\\&=(\hat{\alpha}-\alpha  [\mathbf{E}\mathbf{B}]_{kk})\gamma_k^\frac12{s}_{t,k} - \alpha \sum_{l \neq k}[\mathbf{E}\mathbf{B}]_{kl}\gamma_l^\frac12{s}_{t,l} + n_{t,k} \\
    &=(\hat{\alpha}-\alpha\varepsilon_{kk})\gamma_k^\frac12 {s}_{t,k} - \alpha \sum_{l \neq k}\varepsilon_{kl}\gamma_l^\frac12{s}_{t,l} + n_{t,k},
\end{aligned}
\end{equation}
where $\varepsilon_{kl}=[\mathbf{E}\mathbf{B}]_{kl}$. Note in this equation that channel estimation errors appear both in the additive MUI (second term) as well as uncertainty in the amplitude of symbols (first term in the summation). 

\subsection{MULTIUSER INTERFERENCE}
The estimation error $\mathbf{E}$, as computed in Sec. \ref{sec:channel_estimation}, comes from the estimation of $\mathbf{V}_k$, so we can relate $\mathbf{E}$ to $\mathbf{E}^{(1)}$ and $\mathbf{E}^{(2)}$ in \eqref{eq:H_estimated} as:
\begin{equation}    
\mathbf{E}=(\mathbf{I}_M\otimes\phiM^T)\mathbf{E}^{(2)}+\mathbf{E}^{(1)},
\end{equation}
a matrix whose components are Gaussian, zero-mean, and uncorrelated as proved in \eqref{eq:cov_mat}. To evaluate the variance of $\varepsilon_{kl}$ in \eqref{eq:y_epsilon}, we proceed as follows. 

Take $\mathbf{b}_k$ as the $k$-th column of the precoding matrix $\mathbf{B}$ and compute the expression:
\begin{equation}
\label{eq:analyze_E}
\begin{split}
    \left [ \mathbb{E} \left \{\mathbf{E}^{(2)}\mathbf{b}_k\mathbf{b}_k^H\mathbf{E}^{(2)H}\right \} \right ]_{ij} =\mathbb{E} \left \{\mathbf{e}^{(2)T}_i\mathbf{b}_k\mathbf{b}_k^H\mathbf{e}^{(2)*}_j\right \} \\
    = \mathbb{E} \left \{\mathbf{b}_k^H\mathbf{e}^{(2)*}_i\mathbf{e}^{(2)T}_j\mathbf{b}_k\right \} =\mathbf{b}_k^H\mathbb{E} \left \{\mathbf{e}^{(2)*}_i\mathbf{e}^{(2)T}_j\right \}\mathbf{b}_k \\
    = \left \| \mathbf{b}_k \right \|^2\frac{\sigma^2_{UL}}{P_{UL,j}T_j}\delta_{i-j},
    \end{split}
\end{equation}
where $\delta_{i-j}$ is the Kronecker delta function, and \eqref{eq:cov_mat} is used in the last equality. In this expression, $P_{UL,j}$ is the power transmitted in the UL by the $j$-th user for channel estimation purpose, and $T_j$ is the number of pilot symbols used. Then, we proceed to compute the variance of the MUI $\varepsilon_{kl}$ in equation \eqref{eq:variance_noise} (in next page), where \eqref{eq:analyze_E} is used in equality  $\left(a\right)$, and $\mathbf{f}_k$ is a canonical vector that has a 1 in position $k$ and zeros elsewhere. We have also used that the elements in vector $\phiM$ are unit modulus for a passive RIS.
\begin{figure*}[t]
\normalsize
\begin{equation}
\label{eq:variance_noise}
\begin{aligned}
    \text{var}(\varepsilon_{kl}) &=\mathbf{f}_k^H\mathbb{E} \left \{\mathbf{E}\mathbf{B}\mathbf{f}_l\mathbf{f}_l^H\mathbf{B}^H\mathbf{E}^H \right \}\mathbf{f}_k 
    = \mathbf{f}_k^H\mathbb{E} \left \{\mathbf{E}\mathbf{b}_l\mathbf{b}_l^H \mathbf{E}^H \right \}\mathbf{f}_k \\
    &= \mathbf{f}_k^H\mathbb{E} \left \{ \left ( (\mathbf{I}_M\otimes\phiM^T)\mathbf{E}^{(2)}+\mathbf{E}^{(1)} \right )\mathbf{b}_l\mathbf{b}_l^H\left ( (\mathbf{I}_M\otimes\phiM^T)\mathbf{E}^{(2)}+\mathbf{E}^{(1)} \right )^H \right \}\mathbf{f}_k \\
    &= \mathbf{f}_k^H\left [\left ( \mathbf{I}_M\otimes\phiM^T\right )\mathbb{E} \left \{ \mathbf{E}^{(2)}\mathbf{b}_l\mathbf{b}_l^H\mathbf{E}^{(2)H}\right \}\left ( \mathbf{I}_M\otimes\phiM^* \right )+
    \mathbb{E} \left \{\mathbf{E}^{(1)}\mathbf{b}_l\mathbf{b}_l^H\mathbf{E}^{(1)H}\right \}\right ]\mathbf{f}_k \\&\stackrel{(a)}{=}
\mathbf{f}_k^H \left [ \left ( \mathbf{I}_M\otimes\phiM^T\right )\left \| \mathbf{b}_l \right \|^2\frac{\sigma_{UL}^2}{P_{UL,l}T_l} \left ( \mathbf{I}_M\otimes\phiM^* \right )+\left \| \mathbf{b}_l \right \|^2\frac{\sigma_{UL}^2}{P_{UL,l}T_l}\mathbf{I}_M \right ]\mathbf{f}_k \\
    &=\mathbf{f}_k^H\left [ \left \| \mathbf{b}_l \right \|^2 \frac{\sigma_{UL}^2}{P_{UL,l}T_l}\left ( \left ( \mathbf{I}_M\otimes\phiM^T\phiM^* \right ) + \mathbf{I}_M \right ) \right ]\mathbf{f}_k = \left (1+N_r\right )\left \| \mathbf{b}_l \right \|^2 \frac{\sigma_{UL}^2}{P_{UL,l}T_l}
\end{aligned}
\end{equation}
\hrulefill
\end{figure*}
The MUI plus noise power at user $k$ is therefore obtained as the sum of powers of the second and third terms in     \eqref{eq:y_epsilon}:
\begin{equation}
    \label{eq:sigma_k}
    \sigma_k^2=\alpha^2(1+N_r)\sigma_{UL}^2\sum_{l \neq k} \frac{\gamma_l\left \| \mathbf{b}_l \right \|^2}{P_{UL,l}T_l} + \sigma^2.
\end{equation}

\subsection{AMPLITUDE UNCERTAINTY}
Let us focus on the impact of the first term in equation \eqref{eq:y_epsilon} on the BER. Conditioning on the error observed in the channel for user $k$, we can write the BER as:
\begin{equation}
    P_{e,k} = \int_{-\infty}^{\infty}P_b \left (SNR_k(\varepsilon_{kk})\right ) f(\varepsilon_{kk})d\varepsilon_{kk},
    \label{eq:integral_of_Pb}
\end{equation}
where $P_{b}$ is the bit error probability, which depends on the signal-to-noise ratio $SNR_k(\varepsilon_{kk})=\hat{\alpha}^2|1-\varepsilon_{kk}|^2\gamma_k/\sigma_k^2$. We will consider that $\varepsilon_{kk} \sim {\mathcal{CN}}(0,\sigma_{\epsilon k}^2)$. The expression for $\sigma_{\epsilon k}^2=\text{var}(\epsilon_{kk})$ can be trivially obtained from \eqref{eq:variance_noise}.

\subsection{BER FOR AN A-QAM TRANSMISSION}
\label{subsectionBER-QAM}
For BER evaluation, we assume transmissions using square-shaped $A$-QAM constellations, for which an approximate BER in coherent demodulation is \cite{Proakis2007}:
\begin{equation}
\label{eq:Pb_QAM}
    P_b(SNR) \approx \frac{4}{\log_2A}\left(1-\frac{1}{\sqrt{A}}\right)Q\left(\sqrt{\frac{3\:{SNR}}{A-1}}\right),
\end{equation}
where $Q$ is the Gaussian integral function.  Plugging \eqref{eq:Pb_QAM} in \eqref{eq:integral_of_Pb} and using the improved exponential bound of the $Q$ function described in \cite{ChernoffBoundQ}:
\begin{equation}
    Q(x) \leq \frac{1}{4}\exp(-x^2)+\frac{1}{4}\exp(-x^2/2),
\end{equation}
the BER for user $k$ $P_{e,k}$ is upper bounded as:
\begin{multline}
    P_{e,k} \leq \frac{2(\sqrt{A}-1)}{\log_2A\sqrt{2A(2+\sigma_{\epsilon k}^2 Z_{k}})}\exp \left (-\frac{Z_k}{2+\sigma_{\epsilon k}^2 Z_{k}}\right) \\
    + \frac{2(\sqrt{A}-1)}{\log_2A\sqrt{2A(1+\sigma_{\epsilon k}^2 Z_{k}})}\exp \left (-\frac{Z_k}{1+\sigma_{\epsilon k}^2 Z_{k}}\right),
    \label{eq:Pek}
\end{multline}
where $Z_k=\frac{3}{A-1}\frac{\hat{\alpha}^2\gamma_k}{\sigma_k^2}$. The reader may check that when the channel is perfectly estimated, $\sigma^2_{\epsilon k}=0$ the probability of error follows the improved exponential bound of an $A$-QAM modulation. Otherwise, channel estimation errors induce a BER floor. 

\subsection{ROBUST DESIGN}
From \eqref{eq:sigma_k}, it appears that imperfect channel knowledge may affect every user $k$ differently. If the $k$-th row of $\Ht$, associated to user $k$, has a small norm, then $\left\|\mathbf{b}_k\right\|^2$ will be large and the rest of users will be much affected by MUI. Only in the particular case that the direct channel is negligible compared to reflected channel, and $\Ht_{br}$ and $\Ht_{ru}$ are not rank deficient matrices, the asymmetries in MUI across users may disappear. As a result of optimizing $\alpha$ we have the following properties:

\begin{lemma}
\label{theorem 2}
     If $\Ht_{bu}=\mathbf{0}$, the maximum $\alpha^2$ is attained when the eigenvalues of $\mathbf{\Gamma}^\frac12(\mathbf{H}\mathbf{H}^H)^{-1}\mathbf{\Gamma}^\frac12$ are equal.
\end{lemma}

The proof is in Appendix \ref{Bound2 on alpha}.
Using this lemma, we can derive the following result:

\begin{theorem}
\label{singvalofS}
    If all singular values of $\mathbf{\Gamma}^{-\frac12}\mathbf{H}$ are equal, so are the terms $\gamma_l\left \| \mathbf{b}_l \right \|^2$ for $l=1,\dots,K$.
\end{theorem}

The proof is in Appendix \ref{NormOfColumnsOfB}.
Full-rank in $\Ht_{br}$ and $\Ht_{ru}$ may be achieved even in LOS if the BS and UE are in the near-field of the RIS, which is the situation for the simulations setup defined in section \ref{Results}. Therefore, in scenarios where the direct path is highly attenuated, the optimization procedure has the potential to equalize the MUI in \eqref{eq:sigma_k} across users, as long as the energy spent in the transmission of pilot symbols ${P_{UL,l}T_l}$ is also the same. 

In case the direct channel is not negligible, the impact of channel estimation errors is more relevant and appears through the variances $\sigma_{\epsilon k}^2$ (the amplitude uncertainty) and $\sigma_k^2$ (the MUI plus noise). If interested in a fair transmission across users, the same $P_{e,k}$ for all $k=1,\dots,K$ is achieved by enforcing the same  $Z_k$ and also the same $\sigma_{\epsilon k}^2 Z_{k}$ for all. Consequently all $\sigma_{k}^2/\gamma_k$ and all $\sigma_{\epsilon k}^2$ should be equal.

A robust solution for equal BER across users would design the system in such a way that $\frac{\gamma_l \left \| \mathbf{b}_l \right \|^2}{P_{UL,l}T_l}$ in \eqref{eq:sigma_k} is the same for all users. This entails that each user has to adjust the energy spent in the transmission of its pilot symbols, $P_{UL,l}T_l$, which is not straightforward as $\Ht$ includes the phases of the RIS and, hence, neither $\Ht$ or $\left \| \mathbf{b}_l \right \|^2$ are known before optimization. This difficulty can be circumvented by assuming that consecutive-in-time channel estimates are only slightly changing the values of $\left \| \mathbf{b}_l \right \|^2$. As $\mathbf{B}$ is computed at the BS, it should communicate each UE the amount of energy devoted to pilots.

Finally, unequal per-user received powers can be appointed by choosing $\gamma_k$ (see equation     \eqref{eq:y_model}). Those users with higher $\gamma_k$ should increase the energy of their pilots accordingly so that not to alter MUI.

\section{RESULTS} \label{Results}
In this section, we provide results for the BER of a \ac{ris}-assisted MU-MISO communication with $K\geq2$, with a twofold purpose: firstly, evaluate the two difficult scenarios for MU-MISO transmissions described in the introduction; and secondly, to assess the impact of imperfect channel state information. In all cases, the BER for a 4-QAM transmission is used as a performance indicator.
\subsection{SCENARIO DESCRIPTION}
In a first set of simulations, two users are deployed: UE$_1$ is placed at every position on a $2\:m$ resolution grid of a $120\:m\times 120\:m$ area and a height of $1.5\:m$, while UE$_2$ is placed at a fixed position $[20,35,1.5] \:m$ (the green dot in Fig. \ref{fig:4x2_direct_4qam} through Fig. \ref{fig:4x2_out_ris}). Moreover, we incorporate two obstacles of 10 $m$ wide (marked as black lines in Fig. \ref{fig:4x2_direct_4qam} and Fig. \ref{fig:4x2_nr1000_4qam}) that obstruct the radio propagation with a 10 dB attenuation. $\beta_{bu,k,l}$ in \eqref{eq:direct_channel_su} is calculated using the model in \cite{3gpp.38.901}[Table 7.4.1-1] for a micro scenario with $d_{k,m}\geq10$, considering that both BS and UE antennas are isotropic.
\begin{table}[b]
    \centering
    \caption{System simulation parameters.}
    \begin{tabular}{|l|l|}
    \hline
        \textbf{Parameters} & \textbf{Values} \\ \hline
        Carrier Frequency, $f_c$ & 30 GHz  \\ \hline
        Antenna gains  & $G_t$  = $G_u$  = 3 dBi  \\ \hline
        Transmit power in DL, $P_t$ & 24 dBm  \\ \hline
        Receiver noise power & -94 dBm  \\ \hline
        Channel bandwidth, $B_w$ & 50 MHz\\  \hline
        BS location \& tilt (azim./elev.) & [60,120,10] m, $(\pi,\pi/2)$  \\ \hline
        UE$_1$ height \& UE$_2$ location & 1.5 m, [20,35,1.5] m  \\ \hline
        RIS location \& tilt (azim./elev.) & [0,60,6] m, $(-\pi/2,\pi/2)$  \\ \hline
        Obstacles location \& orientation & [40,80],[80,70], along the $x$-axis      \\ \hline
    \end{tabular}
    \label{system_param}
\end{table}
Following the guidelines in \cite{Ozdogan_IRSmodeling}, we use  \cite{balanis2012advanced}[Example 11-3] to find the channel gains of the compound channel for every reflecting RIS element of dimensions $a\times b$: 
\begin{gather}
    g_{ru,k,i}=\frac{G_u}{4\pi}\frac{ab}{d_{k,i}^2}\text{sinc}^2 (Y)\text{sinc}^2(W), \\
    g_{br,i,l}=\frac{G_t}{4\pi}\frac{ab}{d_{i,l}^2}\cos^2\psi_i, \\
    W = \frac{\pi a}{\lambda_c} \cos\theta_s,\:\:\:
    Y = \frac{\pi a}{\lambda_c} \left ( \sin\psi_i+\sin\theta_s\sin\psi_s\right),
\end{gather}
where $\psi_i$ corresponds to the azimuth angle of the incident wave on the RIS panel, while $\psi_s$ and $\theta_s$ correspond to the azimuth and elevation angles of the scattered wave. The RIS elements are distributed in $N_z=5$ rows and $N_y=N_r/5$ columns. We adopt $a=b=0.5\lambda_c$ and $M=4$ antennas spaced by $0.5\lambda_c$ at the BS. Other parameters are in Table \ref{system_param}.

The Fraunhofer distance, defining the border between near-field and far-field, is given by $d_f=\frac{2D^2}{\lambda},$ where $D$ is the maximum dimension of the radiator. At our carrier frequency, $\lambda=0.01\:m$. For the 4 antennas BS array, $d_f=0.0225\:m$, and for a RIS of $5\times200$ elements, $d_f=198.025\:m$. As per the frequency selectivity of the array apertures, the propagation delay is $0.1\:ns$ for the BS and $6.6\:ns$ for the RIS, well below the $20\:ns$ symbol time. 

To evaluate the benefits of RIS in the relative positions of UE$_1$ and UE$_2$ and the impact of channel estimation errors, we propose to study four relevant configurations: 
\begin{itemize}
    \item NEAR. UE$_2$ is placed at coordinates $[20, 40, 1.5] \:m$, close to the RIS and behind one of the obstacles.
    \item OUT. UE$_2$ is placed at coordinates $[40, 40, 1.5] \:m$, outside the shadowed areas and close to the RIS.
    \item FAR. UE$_2$ is placed at $[90, 40, 1.5] \:m$, behind one of the obstacles and far from the RIS.
    \item NEAR-NS. UE$_2$ is placed at $[20, 35, 1.5] \:m$ and there are no obstacles in the simulation.
\end{itemize}
In all configurations, UE$_1$ is placed at every position of the area, except on top of UE$_2$. In this way, the color maps in Fig. \ref{fig:4x2_direct_4qam} through Fig. \ref{fig:4x2_out_ris} show the BER of both UE$_1$ and UE$_2$ (as BER is the same for both when perfect channel state knowledge), at every possible place of UE$_1$ in the area.

\subsection{BER IN DIRECT AND RIS-ASSISTED TRANSMISSIONS}
In this section we assume perfect channel estimation, so both terminals experience the same SNR and BER as a result of ZF transmission, if $\gamma_1=\gamma_2$ (see \eqref{eq:norm_factor}). 
\subsubsection{RESILIENCE TO OBSTACLES}
We evaluate the BER from \eqref{eq:Pb_QAM} of the 2-users MISO $4$-QAM RIS-assisted transmission in the DL with $N_r=1000$ in the NEAR scenario. To assess the impact of RIS deployment, simulations are conducted with and without RIS. 

\begin{figure}[t]
    \begin{center}
        \includegraphics[scale=0.62]{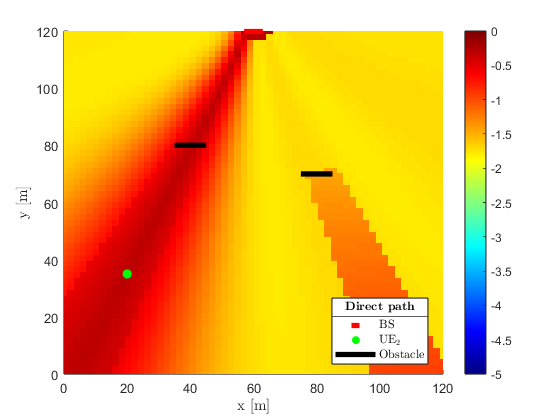}
        \caption{$\log_{10}$(BER) on a coverage area for $4$-QAM MU-MISO transmission (no RIS present) in the NEAR configuration.}
        \label{fig:4x2_direct_4qam}
    \end{center}
\end{figure}

\begin{figure}[t]
    \begin{center}
        \includegraphics[scale=0.62]{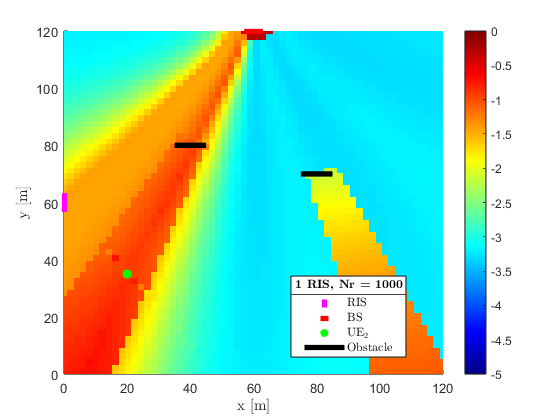}
        \caption{$\log_{10}$(BER) on a coverage area for $4$-QAM with a single RIS, $N_r=1000$ in the NEAR configuration.}
        \label{fig:4x2_nr1000_4qam}
    \end{center}
\end{figure}

\begin{figure}[t]
    \begin{center}
        \includegraphics[scale=0.5]{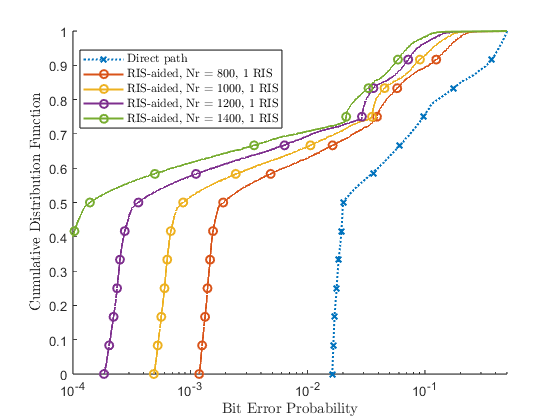}
        \caption{CDF of BER for $4$-QAM with perfect channel knowledge in the NEAR setup. $M=4$ antennas at BS and $N_r$ RIS elements.}
        \label{fig:CDF_4QAM_M4_PE}
    \end{center}
\end{figure}

As shown in Fig. \ref{fig:4x2_direct_4qam} and Fig. \ref{fig:4x2_nr1000_4qam}, UE$_2$ suffers high attenuation of the direct path to the BS. In the absence of RIS in Fig. \ref{fig:4x2_direct_4qam}, the coverage area shows patterns of signal attenuation and fading, influenced by the propagation losses, the obstacles and the alignment of the two users. On its turn, see Fig. \ref{fig:4x2_nr1000_4qam}, the presence of optimized \ac{ris}s enhances the received SNR, significantly reduces BER all over the coverage area, and improves as the number of \ac{ris} elements increases (as can be checked in the CDF of BER in Fig. \ref{fig:CDF_4QAM_M4_PE}).

Note in Fig. \ref{fig:CDF_4QAM_M4_PE} that for each RIS configuration there are two steep slopes at different values. The leftmost slope is due to areas that are not shadowed by obstacles; therefore, the higher $N_r$ is, the lower is the BER. The slope on the top-right part of the plot is due to shadowed areas, and the central BER value (between $2\times10^{-1}$ and $8\times10^{-1}$) is similar for all the range of $N_r$ values, but still well below the BER obtained without RIS (blue curve). Therefore, increasing $N_r$ is more effective when direct path is present, that is, when UE$_1$ is in non-shadowed areas.

\subsubsection{RESILIENCE TO UE ALIGNMENT}
\begin{figure}[t]
\label{figBERNEARNSnoRIS}
    \begin{center}
        \includegraphics[scale=0.42]{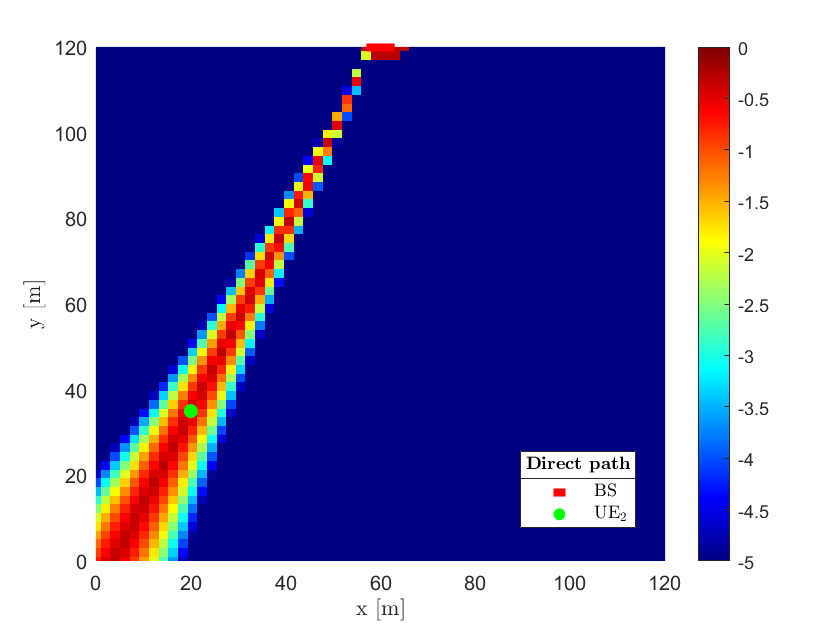}
        \caption{$\log_{10}$(BER) on a coverage area for $4$-QAM MU-MISO transmission (no RIS present) in the NEAR-NS configuration. BER values below $10^{-5}$ have been clipped.}
        \label{fig:4x2_out_dp}
    \end{center}
\end{figure}

\begin{figure}[t]
\label{figBERNEARNSRIS}
    \begin{center}
        \includegraphics[scale=0.42]{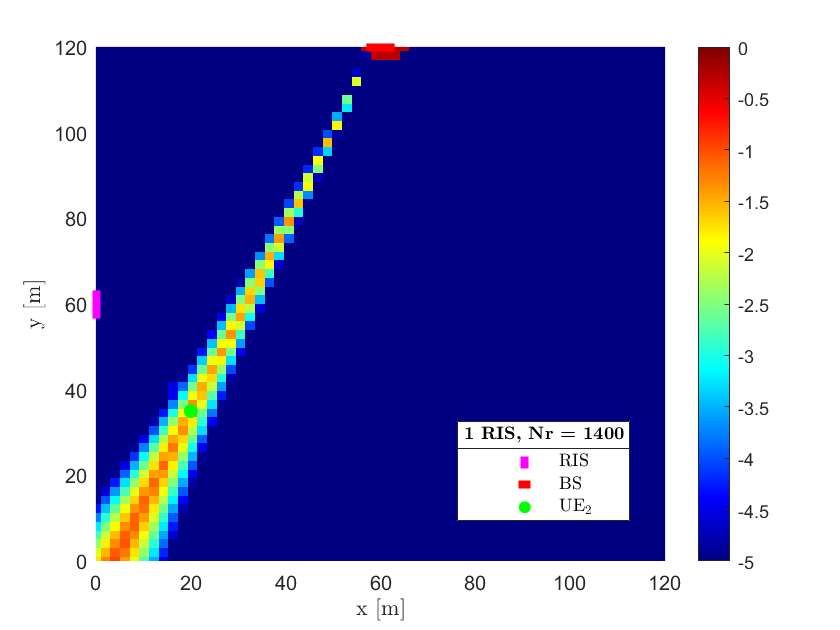}
        \caption{$\log_{10}$(BER) on a coverage area for $4$-QAM with a single RIS, $N_r=1400$ in the NEAR-NS configuration. BER values below $10^{-5}$ have been clipped.}
        \label{fig:4x2_out_ris}
    \end{center}
\end{figure}

We study the NEAR-NS scenario with $N_r= 1400$, where no obstacles are present. Figs. \ref{fig:4x2_out_dp} and \ref{fig:4x2_out_ris} display the BER in the NEAR-NS scenario to assess the role of RIS when users are in the far field and aligned with the BS. Without RIS deployment, the BER experienced by aligned users is significantly higher than for non-aligned positions due to the poor condition of the channel matrix $\Ht$. The use of an optimized RIS mitigates this situation as a result of the additional propagation path that increases the received signal power, leading to a notable reduction in the BER observed.

\subsection{IMPACT OF CHANNEL ESTIMATION ERRORS}
We evaluate the BER over the area when the channel is estimated with errors (see Sec. \ref{sec:channel_unc}). We assume an uplink transmitted power for pilot symbols of $P_{UL}=15$ dBm and $\sigma_{n,UL}^2=\sigma_{n,DL}^2$. We compare the perfect estimation (PE) and imperfect estimation (IE) cases and adopt the same improved Chernoff bound for BER to both PE and IE, on the scenarios described in section \ref{subsectionBER-QAM}.

\begin{figure}[t]
    \begin{center}
        \includegraphics[scale=0.34]{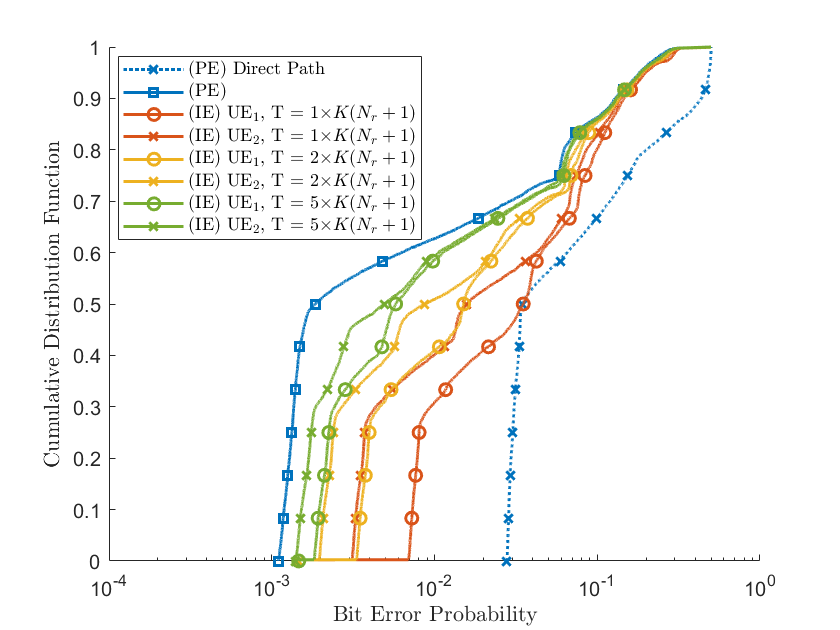}
        \caption{Impact of channel estimation errors in the NEAR setup when increasing the number of pilot symbols $T$. CDF of BER for $K=2$ users, for $4$-QAM, $M=4, N_r=1000$ with perfect channel estimation (PE) and imperfect estimation (IE). BER is computed from the Chernoff bound.}
        \label{fig:CDF_NEAR_PEvsIE_T}
    \end{center}
\end{figure}

\begin{figure}[t]
    \begin{center}
        \includegraphics[scale=0.5]{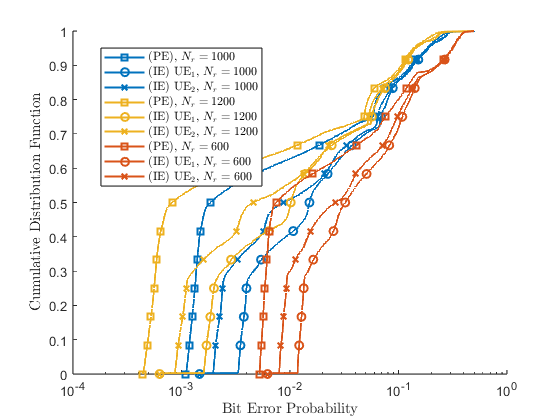}
        \caption{Impact of channel estimation errors in the NEAR configuration when increasing in the number of RIS elements $N_r$. CDF of BER for $4$-QAM, $M=4, T=2(N_r+1)$ with perfect estimation (PE) and imperfect estimation (IE). All BER are computed using the Chernoff bound.}
        \label{fig:NEAR_Nr}
    \end{center}
\end{figure}

Fig. \ref{fig:CDF_NEAR_PEvsIE_T} plots the CDF of BER for the NEAR setup with $N_r=1000$. Note that for the IE case, the two users experience different BER. It shows the detrimental effect of channel estimation error in BER for several values of the number of pilots $T$. It can also be observed that even for the minimum number of pilots $T=K(N_r+1)$, the BER for both users is considerably lower than when no RIS is used (dashed blue line). Interestingly, Fig. \ref{fig:NEAR_Nr} shows that although adding more reflecting elements lowers the BER, it does not yield improved resilience to channel estimation errors.

\subsection{IMPACT OF RELATIVE UE$_1$ AND UE$_2$ POSITIONS}

Table \ref{TableOutageBER} displays outage BER for the four different configurations when perfect estimation is assumed. RIS-aided communication consistently outperforms direct paths across all cases and outage percentages (50\% and 80\%).
Yet, when considering the effect of imperfect estimation (IE), the performance of RIS-aided communication varies significantly depending on the scenario.

For the FAR configuration, the BER for RIS-aided communication with imperfect estimation (UE$_1$ and UE$_2$) is slightly higher compared to both direct paths and RIS-aided communication with perfect estimation: channel estimation errors significantly degrade performance, since UE$_2$ is behind the obstacle and also far from the BS and the RIS, which entails a much lower received SNR. However, the performance degradation in the FAR case could be compensated by using additional RISs elements (not shown).

In contrast, for the NEAR and NEAR NS configurations, RIS-aided communication still outperforms direct transmission, albeit with a slight degradation when channel is estimated with errors. Similarly to NEAR-NS, the OUT configuration provides the lowest BER values, since UE$_2$ is not shadowed in both scenarios.

Among the four studied scenarios, the biggest improvement when using RIS is in the NEAR configuration, where UE$_2$ suffers signal attenuation due to an obstacle near the RIS: at the median value, the BER drops two orders of magnitude with respect to transmissions without RIS ($N_r=1000$ is used). 


\begin{table}[]
\caption{50\% and 85\% outage BER for both users in the four configurations of two user terminal-RIS-obstacles, for perfect (PE) and imperfect (IP) channel estimation. In all cases the total energy devoted to pilot symbol transmissions is $E_p=P_tK(N_r+1)$, where $K=2$ users and $N_r=1000$. Recall that under PE, both users experience the same SNR and BER. For the IE case, each user's BER is different, as follows from \eqref{eq:Pek}.}
\label{TableOutageBER}
\centering
\begin{tabular}{|ccl|}
\hline
\multicolumn{1}{|c|}{}                                              & \multicolumn{1}{c|}{\textbf{50\%}} & \multicolumn{1}{c|}{\textbf{85\%}} \\ \hline
\multicolumn{3}{|c|}{\textit{FAR configuration}}                                                                                                          \\ \hline
\multicolumn{1}{|c|}{\textbf{Direct path (PE)}}                     & \multicolumn{1}{c|}{2.34E-02}     & 2.64E-01                          \\ \hline
\multicolumn{1}{|c|}{\textbf{RIS-aided (PE)}}                       & \multicolumn{1}{c|}{1.76E-02}     & 1.74E-01                          \\ \hline
\multicolumn{1}{|c|}{\textbf{RIS-aided (IE) UE$_1$}} & \multicolumn{1}{c|}{5.91E-02}     & 2.12E-01                          \\ \hline
\multicolumn{1}{|c|}{\textbf{RIS-aided (IE) UE$_2$}} & \multicolumn{1}{c|}{4.46E-02}     & 2.15E-01                          \\ \hline
\multicolumn{3}{|c|}{\textit{NEAR configuration}}                                                                                                         \\ \hline
\multicolumn{1}{|c|}{\textbf{Direct path (PE)}}                     & \multicolumn{1}{c|}{3.45E-02}     & 3.01E-01                          \\ \hline
\multicolumn{1}{|c|}{\textbf{RIS-aided (PE)}}                       & \multicolumn{1}{c|}{1.85E-03}     & 8.56E-02                          \\ \hline
\multicolumn{1}{|c|}{\textbf{RIS-aided (IE) UE$_1$}} & \multicolumn{1}{c|}{1.51E-02}     & 9.69E-02                          \\ \hline
\multicolumn{1}{|c|}{\textbf{RIS-aided (IE) UE$_2$}} & \multicolumn{1}{c|}{8.73E-03}     & 9.40E-02                          \\ \hline
\multicolumn{3}{|c|}{\textit{NEAR NS configuration}}                                                                                                      \\ \hline
\multicolumn{1}{|c|}{\textbf{Direct path (PE)}}                     & \multicolumn{1}{l|}{2.42E-50}     & 5.24E-10                          \\ \hline
\multicolumn{1}{|c|}{\textbf{RIS-aided (PE)}}                       & \multicolumn{1}{l|}{7.61E-53}     & 1.06E-12                          \\ \hline
\multicolumn{1}{|c|}{\textbf{RIS-aided (IE) UE$_1$}} & \multicolumn{1}{l|}{1.24E-39}     & 1.64E-08                          \\ \hline
\multicolumn{1}{|c|}{\textbf{RIS-aided (IE) UE$_2$}} & \multicolumn{1}{l|}{1.50E-41}     & 1.20E-08                          \\ \hline
\multicolumn{3}{|c|}{\textit{OUT configuration}}                                                                                                          \\ \hline
\multicolumn{1}{|c|}{\textbf{Direct path (PE)}}                     & \multicolumn{1}{l|}{4.60E-53}     & 5.77E-02                          \\ \hline
\multicolumn{1}{|c|}{\textbf{RIS-aided (PE)}}                       & \multicolumn{1}{l|}{3.10E-55}     & 9.84E-03                          \\ \hline
\multicolumn{1}{|c|}{\textbf{RIS-aided (IE) UE$_1$}} & \multicolumn{1}{l|}{3.02E-28}     & 1.43E-02                          \\ \hline
\multicolumn{1}{|c|}{\textbf{RIS-aided (IE) UE$_2$}} & \multicolumn{1}{l|}{6.70E-29}     & 1.54E-02                          \\ \hline
\end{tabular}
\end{table}

The comparison across the four configurations highlights is done also in terms of achievable rate, using $C_k = \log_2(1+SNR_k)$ 
with $SNR_k = P\alpha^2 (1+\sigma_{ek}^2)/\sigma_{kk}^2$,
in Table \ref{TableOutageC}, showing a strong correlation with BER. 

Configurations FAR and NEAR exhibit relatively high BER values, which translate into very low spectral efficiency and thus limited capacity. In contrast, NEAR-NS and OUT achieve much lower BERs, enabling significantly higher rates. The minimum observed capacity across the scenarios is approximately 2.5 bps/Hz, while the maximum reaches around 8.5 bps/Hz, at 85\% outage probability over the area.
\begin{table}[]
\caption{50\% and 85\% outage achievable rate for both users in the four configurations of two user terminal-RIS-obstacles, for perfect (PE) and imperfect (IP) channel estimation. In all cases the total energy devoted to pilot symbol transmissions is $E_p=P_tK(N_r+1)$, where $K=2$ users and $N_r=1000$. Recall that under PE, both users experience the same SNR.}
\centering
\begin{tabular}{|ccc|}
\hline
\multicolumn{1}{|c|}{}                               & \multicolumn{1}{c|}{\textbf{50\%}} & \textbf{85\%} \\ \hline
\multicolumn{3}{|c|}{\textit{FAR configuration}}                                                          \\ \hline
\multicolumn{1}{|c|}{\textbf{Direct path (PE)}}      & \multicolumn{1}{c|}{2.56}          & 2.70          \\ \hline
\multicolumn{1}{|c|}{\textbf{RIS-aided (PE)}}        & \multicolumn{1}{c|}{2.68}          & 2.79          \\ \hline
\multicolumn{1}{|c|}{\textbf{RIS-aided (IE) UE$_1$}} & \multicolumn{1}{c|}{1.90}          & 2.31          \\ \hline
\multicolumn{1}{|c|}{\textbf{RIS-aided (IE) UE$_2$}} & \multicolumn{1}{c|}{2.52}          & 2.91          \\ \hline
\multicolumn{3}{|c|}{\textit{NEAR configuration}}                                                         \\ \hline
\multicolumn{1}{|c|}{\textbf{Direct path (PE)}}      & \multicolumn{1}{c|}{2.38}          & 2.46          \\ \hline
\multicolumn{1}{|c|}{\textbf{RIS-aided (PE)}}        & \multicolumn{1}{c|}{3.44}          & 3.54          \\ \hline
\multicolumn{1}{|c|}{\textbf{RIS-aided (IE) UE$_1$}} & \multicolumn{1}{c|}{2.51}          & 3.02          \\ \hline
\multicolumn{1}{|c|}{\textbf{RIS-aided (IE) UE$_2$}} & \multicolumn{1}{c|}{3.16}          & 3.61          \\ \hline
\multicolumn{3}{|c|}{\textit{NEAR NS configuration}}                                                      \\ \hline
\multicolumn{1}{|c|}{\textbf{Direct path (PE)}}      & \multicolumn{1}{c|}{7.82}          & 8.50          \\ \hline
\multicolumn{1}{|c|}{\textbf{RIS-aided (PE)}}        & \multicolumn{1}{c|}{7.90}          & 8.61          \\ \hline
\multicolumn{1}{|c|}{\textbf{RIS-aided (IE) UE$_1$}} & \multicolumn{1}{c|}{7.33}          & 8.12          \\ \hline
\multicolumn{1}{|c|}{\textbf{RIS-aided (IE) UE$_2$}} & \multicolumn{1}{c|}{7.57}          & 8.39          \\ \hline
\multicolumn{3}{|c|}{\textit{OUT configuration}}                                                          \\ \hline
\multicolumn{1}{|c|}{\textbf{Direct path (PE)}}      & \multicolumn{1}{c|}{7.90}          & 8.82          \\ \hline
\multicolumn{1}{|c|}{\textbf{RIS-aided (PE)}}        & \multicolumn{1}{c|}{7.96}          & 8.84          \\ \hline
\multicolumn{1}{|c|}{\textbf{RIS-aided (IE) UE$_1$}} & \multicolumn{1}{c|}{7.48}          & 8.42          \\ \hline
\multicolumn{1}{|c|}{\textbf{RIS-aided (IE) UE$_2$}} & \multicolumn{1}{c|}{6.97}          & 8.54          \\ \hline
\end{tabular}
\label{TableOutageC}
\end{table}

\subsection{BER WHEN TRANSMITTING TO MORE THAN 2 UE}
\begin{figure}[t]
    \begin{center}
        \includegraphics[scale=0.342]{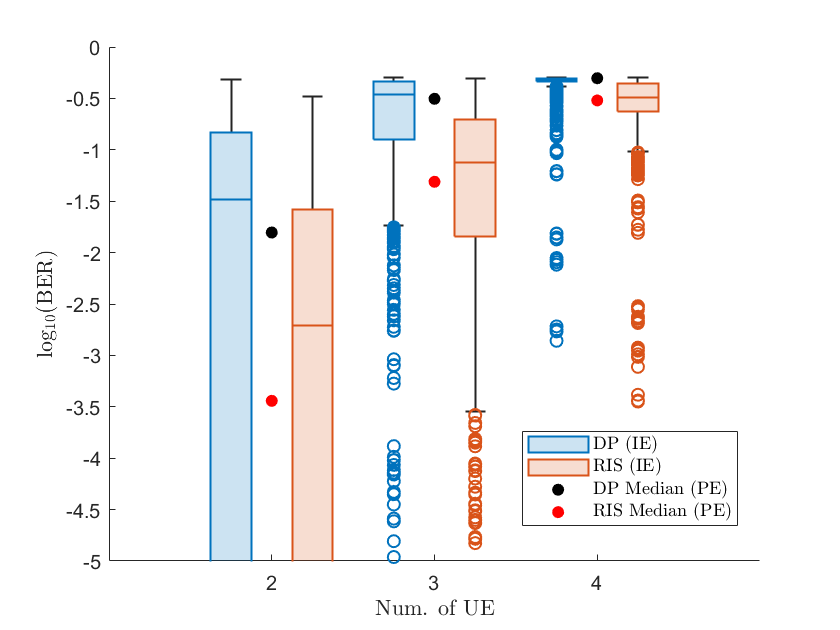}
        \caption{Impact of channel estimation errors for $K=2,3,4$ simultaneous UEs. BER is displayed for $4$-QAM, $M=6, N_r=1500$ with perfect estimation (PE) and imperfect estimation (IE), for direct path (DP) and RIS-assisted transmissions}
        \label{fig:Compare_diff_numUEs}
    \end{center}
\end{figure}

\begin{figure}[t]
    \begin{center}
        \includegraphics[scale=0.5]{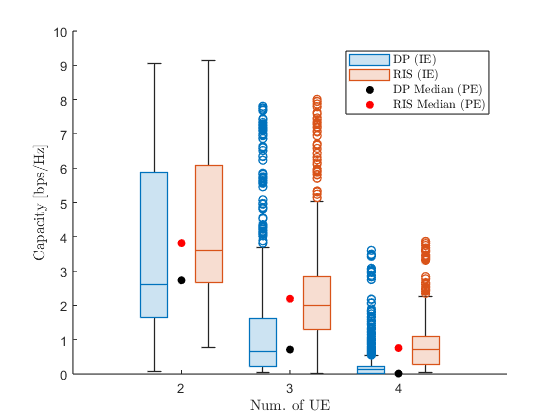}
        \caption{Impact of channel estimation errors for $K=2,3,4$ simultaneous UEs. Capacity is displayed for $M=6, N_r=1500$ with perfect estimation (PE) and imperfect estimation (IE), for direct path (DP) and RIS-assisted transmissions}
        \label{fig:Compare_diff_numUEs_C}
    \end{center}
\end{figure}

To better understand the impact of RIS assistance under varying user loads, we introduce a scenario with $K=\{2,3,4\}$. We simulate a total of $1200/K$ positions per UE with $M=6, N_r=1500$ and $T = 2\times K \times (N_r+1)$, and $K=2,3,4$. Fig. \ref{fig:Compare_diff_numUEs} shows the box plot of BER for the realizations obtained when two obstacles are placed as shown in Fig. \ref{fig:4x2_nr1000_4qam}.

Perfect channel knowledge is assumed. As anticipated, the BER tends to increase with the number of active UEs engaged in multiuser transmission. Fig. \ref{fig:Compare_diff_numUEs} clearly illustrates this upward trend as moving from 2 to 4 UEs across all $1200/K$ UE positions. It may be noted that the assistance of RIS introduces a notable reduction in BER. This improvement becomes more significant as the number of UEs increases.

Similar conclusions regarding the achievable rate can be extracted from the boxplot in Fig. \ref{fig:Compare_diff_numUEs_C}.

Additionally, Table \ref{tab:DPvsRIS_diffUEs} compares the number of runs where a minimum BER threshold of 0.01 is achieved. Four configurations are shown: Direct Path (DP), RIS-assisted transmission, and their respective variants with obstacles. RIS configurations consistently outperform DP setups, e.g. when considering obstacles and 2 UEs, RIS-assisted transmissions achieve a BER below threshold in 776 runs, compared to the 468 runs for DP.

\begin{table}[]
\caption{Number of runs where UEs achieve a BER below $0.01$ across different transmission schemes, for $M = 6, K=\{2,3,4\}, T = K \times (N_r+1)$ and $1200/K$ position per UE. The total number of runs under test is 600, 400 and 300 respectively. $N_r=1200$}
\centering
\begin{tabular}{c|c|c|c|}
\cline{2-4}
                                       & \textbf{2 UEs} & \textbf{3 UEs} & \textbf{4 UEs} \\ \hline
\multicolumn{1}{|c|}{\textbf{DP with obstacles}}      & 468             & 91             & 8              \\ \hline
\multicolumn{1}{|c|}{\textbf{RIS with obstacles}}     & 776             & 259             & 20              \\ \hline
\multicolumn{1}{|c|}{\textbf{DP No obstacles}}  & 1077            & 565             & 60              \\ \hline
\multicolumn{1}{|c|}{\textbf{RIS No obstacles}} & 1139            & 702             & 102             \\ \hline
\end{tabular}

\label{tab:DPvsRIS_diffUEs}
\end{table}

\section{CONCLUSIONS}\label{conclusions}
The paper delves into the advantages of utilizing RIS-aided in ZF-precoded MU-MISO wireless communication, focusing on enhancing coverage, macrodiversity, and resilience to user's relative position, measured in terms of reduction of BER for all UEs in various configurations. Complex matrix calculus has been used as a general framework to define a gradient algorithm for the optimization of RIS that maximizes the received signal power at UEs. The analysis reveals that the use of RIS improves the performance of direct transmission ZF precoders in the four relevant scenarios described. RIS also improves the ability of the ZF precoder to serve multiple users. 

Future work will concentrate on extending the proposed framework by incorporating other precoding strategies, such as MMSE, hybrid, and multi-stage designs, to better capture the trade-offs between complexity and performance in RIS-assisted systems. In parallel, advanced CSI estimation methods are crucial to address practical deployment challenges with many RIS elements. Also, adaptive and learning-based optimization techniques which will enable dynamic adjustment to user mobility and channel variations, thus enhancing robustness in realistic scenarios. Finally, exploring more advanced RIS architectures (BD-RIS, STAR-RIS) is also aimed.

\begin{appendices}
\section{PROOF OF THEOREM 1}\label{Bound1 on alpha}
We depart from the following property: if $\mathbf{P}$ and $\mathbf{Q}$ are positive semidefinite matrices, then $\mathrm{Tr}\left\{\mathbf{P}\mathbf{Q}\right\}\geq 0$. By taking $\mathbf{P}=\left( \mathbf{H}\mathbf{H}^{H} \right )^{-1}$ and $\mathbf{Q}=\mathbf{\Gamma}-\gamma_{min}\mathbf{I}$, where $\gamma_{min}$ is the minimum in the diagonal of $\mathbf{\Gamma}$, we can write from  \eqref{eq:norm_factor}: 
\begin{equation}
\begin{aligned}
P\alpha^{-2}=&\:\mathrm{Tr}\left\{\left( \mathbf{H}\mathbf{H}^{H} \right )^{-1} \mathbf{\Gamma}\right\}\geq \gamma_{min}\mathrm{Tr}\left\{\left( \mathbf{H}\mathbf{H}^{H} \right )^{-1}\right\}\\=&\:\gamma_{min}\sum_{i}\frac{1}{\lambda_i}\geq\frac{\gamma_{min}}{\lambda_{min}},
    \end{aligned}
\end{equation}
where $\lambda_{i}$ is an eigenvalue of $\mathbf{H}\mathbf{H}^H$ and $\lambda_{min}$ is the least one.

\section{PROOF OF LEMMA 1}\label{Bound2 on alpha}
Let us first prove that if $\Ht=\Hru\PhiM\Hbr$, the sum of eigenvalues of $\mathbf{\Gamma}^{-\frac12}\Ht\Ht^H\mathbf{\Gamma}^{-\frac12}$ does not depend on $\PhiM$:
\begin{equation}
\label{SumOfEigenvalues}
\begin{aligned}
\mathrm{Tr}\left\{\Ht\Ht^H\mathbf{\Gamma}^{-1}\right\}&=\mathrm{Tr}\left\{\Hru\PhiM\Hbr\Hbr^H\PhiM^H\Hru^H\mathbf{\Gamma}^{-1}\right\}\\&=\mathrm{Tr}\left\{\Hru\PhiM\mathbf{U}\mathbf{\Lambda}_{br}\mathbf{U}^H\PhiM^H\Hru^H\mathbf{\Gamma}^{-1}\right\}\\&=\sum_i\frac{\lambda_{br,i}}{{\gamma_i}}\mathbf{h}_{ru,i}^T\PhiM\mathbf{U}\mathbf{U}^H\PhiM^H\mathbf{h}_{ru,i}\\&=\sum_i\frac{\lambda_{br,i}}{\gamma_i}\mathbf{h}_{ru,i}^T\mathbf{h}_{ru,i},
\end{aligned}
\end{equation}
where $\mathbf{h}_{ru,i}$ is the $i$-th column of $\Hru$. Using the singular value decomposition $\mathbf{\Gamma}^{-\frac12}\mathbf{H}=\mathbf{U\Sigma V}^H$ we can write     \eqref{eq:norm_factor} in terms of the harmonic mean of $\sigma_i^2$, the elements in the diagonal of $\mathbf{\Sigma}^2$:
\begin{equation} 
\begin{aligned}
    \label{inv_of_trace}
        \mathrm{Tr}\left\{\left( \mathbf{H}\mathbf{H}^{H} \right )^{-1}\mathbf{\Gamma}\right\}^{-1}=\left(\sum_{i}\frac{1}{\sigma_i^2}\right)^{-1}
\end{aligned}
\end{equation}
The harmonic mean is upper bounded by the geometric mean, which on its turn is maximized by the arithmetic mean. The maximum is attained when all $\sigma_i^2$ are equal, subject to a fixed value of $\sum_i\sigma_i^2$. Note that from \eqref{SumOfEigenvalues} the sum is fixed as it does not depend on the reflection coefficients of the RIS. Therefore, the geometric mean is maximized when all $\sigma_i^2$ are equal.

\section{PROOF OF THEOREM 2}\label{NormOfColumnsOfB}
The proof proceeds with the SVD decomposition $\mathbf{\Gamma}^{-\frac12}\mathbf{H}=\mathbf{U\Sigma V}^H$ as:
\begin{equation}
    \begin{aligned}
        \gamma_l\left \| \mathbf{b}_l \right \|^2&=\text{diag}(\mathbf{\Gamma}^\frac12\mathbf{B}^H\mathbf{B}\mathbf{\Gamma}^\frac12)_{l,l}\\&=\text{diag}(\mathbf{\Gamma}^\frac12(\mathbf{H}\mathbf{H}^H)^{-1}\mathbf{\Gamma}^\frac12)_{l,l}\\&=\text{diag}(\mathbf{U}\mathbf{\Sigma}^{-2}\mathbf{U}^H)_{l,l}=\sigma^{-2}\text{diag}(\mathbf{U}\mathbf{U}^H)_{l,l}\\&=\sigma^{-2},
    \end{aligned}
\end{equation}
where in the last step we use lemma 1 and assume unit norm eigenvectors.

\section{PROOF OF GRADIENT EXPRESSIONS}\label{gradient expression}
\subsection{DIFFERENTIAL OF A COMPLEX MATRIX FUNCTION}
Take a complex matrix function of complex matrix variables $\mathbf{J}=\mathbf{G}(\mathbf{Z},\mathbf{Z}^*)$ where in the argument $\mathbf{Z} = \mathbf{X} + j\mathbf{Y}$  the real part $\mathbf{X}$ and imaginary part $\mathbf{Y}$ are independent variables. Under this condition it can be proved that $\mathbf{Z}$ and $\mathbf{Z}^*$ are independent variables \cite{Hjrungnes2011ComplexValuedMD}.
A procedure to find the differentials of $\mathbf{J}$ is:
\begin{equation}
    \label{eqAnn:1.1}
    d\mathbf{J}=\text{First order terms}(\mathbf{G}(\mathbf{Z}+d\mathbf{Z},\mathbf{Z}^*+d\mathbf{Z}^*)-\mathbf{G}(\mathbf{Z},\mathbf{Z}^*)).
\end{equation}
Using the $\text{vec}(\cdot)$ operator (that reshapes a matrix into a column vector) on the differential, we can read out the derivatives $\mathcal{D}_Z\mathbf{G}$ and $\mathcal{D}_{Z^*}\mathbf{G}$ as: 
\begin{equation}
    \label{eqAnn:1.2}
d\text{vec}(\mathbf{J})=(\mathcal{D}_\mathbf{Z}\mathbf{G})d\text{vec}(\mathbf{Z})+(\mathcal{D}_\mathbf{Z^*}\mathbf{G})d\text{vec}(\mathbf{Z}^*\ ),
\end{equation}
where the terms $\mathcal{D}_\mathbf{Z}\mathbf{G}$ and $\mathcal{D}_{\mathbf{Z}^*}\mathbf{G}$ as are the Jacobian matrices of $\mathbf{G}$ with respect to $\mathbf{Z}$ and $\mathbf{Z}^*$ respectively. Writing $d\text{vec}\left(\mathbf{J}\right)$ in the form of \eqref{eqAnn:1.2} requires some mathematical elaboration.
\subsection{CHAIN RULE}
Define the composite function:
\begin{equation}
    \label{eqAnn:2.1}
    \mathbf{J}=\mathbf{G}\left(\mathbf{F}\left(\mathbf{Z},\mathbf{Z}^\ast\right),\mathbf{F}^\ast\left(\mathbf{Z},\mathbf{Z}^\ast\right)\right).
\end{equation} The derivatives can be extracted from the expression:
\begin{equation}
    \label{eqAnn:3}
    d\text{vec}\left(\mathbf{J}\right)=\left(\mathcal{D}_\mathbf{F}\mathbf{G}\right)d\text{vec}\left(\mathbf{F}\right)+\left(\mathcal{D}_{\mathbf{F}^\ast}\mathbf{G}\right)d\text{vec}\left(\mathbf{F}^\ast\right),
\end{equation}
and the complex differentials of $\text{vec}\left(\mathbf{F}\right)$ and $\text{vec}\left(\mathbf{F}^\ast\right)$ are:
\begin{equation}  
\label{eqAnn:2.3}
d\text{vec}\left(\mathbf{F}\right)=\left(\mathcal{D}_\mathbf{Z}\mathbf{F}\right)d\text{vec}\left(\mathbf{Z}\right)+\left(\mathcal{D}_{\mathbf{Z}^\ast}\mathbf{F}\right)d\text{vec}\left(\mathbf{Z}^\ast\right),
\end{equation}
\begin{equation}
\label{eqAnn:2.4}
d\text{vec}\left(\mathbf{F}^\ast\right)=\left(\mathcal{D}_\mathbf{Z}\mathbf{F}^\ast\right)d\text{vec}\left(\mathbf{Z}\right)+\left(\mathcal{D}_{\mathbf{Z}^\ast}\mathbf{F}^\ast\right)d\text{vec}\left(\mathbf{Z}^\ast\right).
\end{equation}
By replacing in the upper expression, we obtain the chain rule for the derivative of a composite function:
\begin{multline}
    \label{eqAnn:2.5}
d\text{vec}\left(\mathbf{J}\right) = \left(\left( \mathcal{D}_{\mathbf{F}}\mathbf{G}\right) \left(\mathcal{D}_{\mathbf{Z}} \mathbf{F}\right) + \left(\mathcal{D}_{\mathbf{F}^*}\mathbf{G}\right)\left(\mathcal{D}_{\mathbf{Z}}\mathbf{F}^* \right)\right)d\text{vec}\left(\mathbf{Z}\right)\\
+\left (\left(\mathcal{D}_{\mathbf{F}}\mathbf{G}\right) \left(\mathcal{D}_{\mathbf{Z}^*} \mathbf{F}\right) + \left(\mathcal{D}_{\mathbf{F}^*}\mathbf{G}\right)\left(\mathcal{D}_{\mathbf{Z}^*}\mathbf{F}^* \right)\right)d\text{vec}(\mathbf{Z}^*) \\
=\mathcal{D}_{\mathbf{Z}}\mathbf{J}d\text{vec}(\mathbf{Z}) + \mathcal{D}_{\mathbf{Z}^*}\mathbf{J}d\text{vec}(\mathbf{Z}^*).
\end{multline}
Clearly, the procedure can be iterated to obtain expressions for the composition of several functions. For the composition of four functions:
\begin{equation}
\begin{aligned}
\mathbf{J}=&\mathbf{G}(\mathbf{F}(\mathbf{E}(\mathbf{H}(\PhiM),\mathbf{H}^\ast(\phi)),\mathbf{E}^\ast(\mathbf{H}(\PhiM),\mathbf{H}^\ast(\PhiM))),\\&
\mathbf{F}^\ast(\mathbf{E}(\mathbf{H}(\PhiM),\mathbf{H}^\ast(\PhiM)),\mathbf{E}^\ast(\mathbf{H}(\PhiM),\mathbf{H}^\ast(\PhiM)))),
\end{aligned}
\end{equation}
we obtain \eqref{eqAnn:dvecJ} in the next page.

\begin{figure*}[t]
\normalsize
\begin{multline}
\label{eqAnn:dvecJ}
    d\text{vec}(\mathbf{J})=\\
    \left ( \begin{matrix}
\left (\mathcal{D}_\mathbf{F}\mathbf{G} \right )\left(\left ( \mathcal{D}_\mathbf{E}\mathbf{F} \right )\left ( \mathcal{D}_\mathbf{H}\mathbf{E} \right )+\left ( \mathcal{D}_{\mathbf{E}^*}\mathbf{F} \right )\left ( \mathcal{D}_\mathbf{H}\mathbf{E}^* \right )\right) + 
\left(\mathcal{D}_{\mathbf{F}^*}\mathbf{G} \right )\left (\left( \mathcal{D}_\mathbf{E}\mathbf{F}^* \right )\left ( \mathcal{D}_\mathbf{H}\mathbf{E} \right )+\left ( \mathcal{D}_{\mathbf{E}^*}\mathbf{F}^* \right )\left ( \mathcal{D}_\mathbf{H}\mathbf{E}^* \right )\right) 
\end{matrix} \right )d\text{vec}\left (  \mathbf{H}\right )\\
+\left ( \begin{matrix}
\left (\mathcal{D}_{\mathbf{F}}\mathbf{G} \right )\left(\left ( \mathcal{D}_\mathbf{E}\mathbf{F}^* \right )\left ( \mathcal{D}_\mathbf{H}\mathbf{E} \right )+\left ( \mathcal{D}_{\mathbf{E}^*}\mathbf{F} \right )\left ( \mathcal{D}_{\mathbf{H}^*}\mathbf{E}^* \right )\right )+
\left (\mathcal{D}_{\mathbf{F}^*}\mathbf{G} \right )\left(\left ( \mathcal{D}_\mathbf{E}\mathbf{F}^* \right )\left ( \mathcal{D}_{\mathbf{H}^*}\mathbf{E} \right )+\left ( \mathcal{D}_{\mathbf{E}^*}\mathbf{F}^* \right )\left ( \mathcal{D}_{\mathbf{H}^*}\mathbf{E}^* \right )\right )
\end{matrix} \right )d\text{vec}\left (\mathbf{H}^*\right )\\
=\left(\mathcal{D}_{\mathbf{H}}\mathbf{J}\right)d\text{vec}\left(\mathbf{H}\right)+\left(\mathcal{D}_{\mathbf{H}^*}\mathbf{J}\right)d\text{vec}\left(\mathbf{H}^*\right).
\end{multline}
\hrulefill
\end{figure*}

\subsection{COMPONENT FUNCTIONS OF $\alpha$ AND THEIR DIFFERENTIALS}
\label{AnnSec3}
The function to be minimized is the scalar  $P\alpha^{-2}$ that can be decomposed as:
\begin{equation}
\label{eqAnn:J}
    J=\Tr\left \{ \left ( \Ht\Ht^H \right )^{-1}\mathbf{\Gamma} \right \}=g\left(\mathbf{F}\left(\mathbf{E}\left(\mathbf{H}(\PhiM),\mathbf{H}^*(\PhiM)\right)\right)\mathbf{\Gamma}\right),
\end{equation}
where $g$ is the trace function, $\mathbf{F}$ is the inverse of a matrix, $\mathbf{E}=\Ht\Ht^H$, and $\Ht$ is defined in \eqref{eq:channel_model_RIS}. The derivatives are:
\begin{equation}
    dg=d\Tr\left\{\mathbf{F}\mathbf{\Gamma}\right\}=\Tr\left\{d\mathbf{F}\mathbf{\Gamma}\right\}=\text{vec}\left(\mathbf{\Gamma}\right)^Td\text{vec}\left(\mathbf{F}\right).
\end{equation}
If $\mathbf{E}$ is invertible, then:
\begin{equation}
    d\mathbf{F}=d\mathbf{E}^{-1}=-\mathbf{E}^{-1}(d\mathbf{E})\mathbf{E}^{-1},
\end{equation}
By applying the $\text{vec}(\cdot)$ operator and the identity $\text{vec}\left(\mathbf{R}\mathbf{A}\mathbf{S}\right)=\left(\mathbf{S}^T\otimes\mathbf{R}\right)\text{vec}(\mathbf{A})$, we can read out the derivative:
\begin{gather}
    d\text{vec}(\mathbf{E}^{-1})=-(\mathbf{E}^{-T}\otimes\mathbf{E}^{-1})d\text{vec}(\mathbf{E}), \\
    d\text{vec}(\mathbf{E}^{-1*})=-(\mathbf{E}^{-H}\otimes\mathbf{E}^*)d\text{vec}(\mathbf{E}^*).
\end{gather}
Considering that $\mathbf{E}=\mathbf{HH}^H$,
\begin{equation}
    d\text{vec}(\mathbf{E})=(\mathbf{H}^*\otimes\mathbf{I}_N)d\text{vec}(\mathbf{H})+(\mathbf{I}_N\otimes\mathbf{H})\mathbf{K}d\text{vec}(\mathbf{H}^*),
\end{equation}
where $\mathbf{K}$ is the commutation matrix that relates $\text{vec}(\mathbf{H}^T)$ and $\text{vec}(\mathbf{H})$ through $\text{vec}(\mathbf{H}^T)=\mathbf{K}\text{vec}(\mathbf{H})$ and has the property $(\mathbf{A}\otimes\mathbf{B})=\mathbf{K}(\mathbf{B}\otimes\mathbf{A})\mathbf{K}$.
Now, using  \eqref{eq:channel_model_RIS},\eqref{eq:compund_channel} and the vec(·) operator we can express the differential of $\Ht$ as:
\begin{equation}
    d\text{vec}(\mathbf{H})=(\Ht_{br}^T\otimes\Ht_{ru})d\text{vec}(\PhiM).
\end{equation}
From here, the differential of the RIS scatter matrix depends on the nature of the RIS. For diagonal RIS $\PhiM=\text{diag}(\pmb{\varphi})=\text{diag}(e^{j\pmb{\phi}})$
and the composite channel matrix is:
\begin{equation}
    \mathbf{H}=\Hbu+\Hru \PhiM\Hbr=\Hbu+\sum_{i=1}^{N_r}\varphi_i\mathbf{h}_{ru,i}\mathbf{h}_{br,i}^T,
\end{equation}
where $\mathbf{h}_{ru,i}$ and $\mathbf{h}_{br,i}^T$ are the $i$-th column of matrix $\Hru$ and the row $i^{th}$ of matrix $\Hbr$, respectively. For the computation of gradients in beyond-diagonal RIS, you may refer to \cite{BDRISMIMO}. From the linearity of the differential operator we can write:
\begin{gather} d\mathbf{H}=\sum_{i=1}^{N_r}d\varphi_i\mathbf{h}_{ru,i}\mathbf{h}_{br,i}^T,\\
\label{dvec(H)}
d\text{vec}(\mathbf{H})=\sum_{i=1}^{N_r}d\varphi_ivec(\mathbf{h}_{ru,i}\mathbf{h}_{br,i}^T)=\mathbf{H}_{12}d\pmb{\varphi},
\end{gather}
where $\mathbf{H}_{12}=\begin{bmatrix}
\text{vec}(\mathbf{h}_{ru,1}\mathbf{h}_{br,1}^T) & \cdots & \text{vec}(\mathbf{h}_{ru,N_r}\mathbf{h}_{br,N_r}^T)
\end{bmatrix}\in \mathbb{C}^{MN\times N_r}$.
Since $\pmb{\varphi}=e^{j\pmb{\phi}}$, then $d\pmb{\varphi}=j\textup{diag}(\pmb{\varphi})d\pmb{\phi}$.
\subsection{GRADIENT EXPRESSION}
By introducing these expressions in \eqref{eqAnn:dvecJ}:
\begin{equation}
\begin{split}
    dJ&=(\mathcal{D}_\mathbf{H}J)d\text{vec}(\mathbf{H})+(\mathcal{D}_{\mathbf{H}^*}J)d\text{vec}(\mathbf{H}^*)\\
    &= (\mathcal{D}_\mathbf{H}J)\mathbf{H}_{12}d\pmb{\varphi}+(\mathcal{D}_{\mathbf{H}^*}J)\mathbf{H}_{12}^*d\pmb{\varphi}^*.
\end{split}
\end{equation}
So, the derivatives can be read out as:
\begin{equation}
\label{eqAnn:DphiJ}
    \mathcal{D}_{\pmb{\varphi}}J=(\mathcal{D}_{\mathbf{H}}J)\mathbf{H}_{12},\:\:\:\:\mathcal{D}_{\pmb{\varphi}^*}J=(\mathcal{D}_{\mathbf{H}^*}J)\mathbf{H}_{12}^*.
\end{equation}
Replacing the derivatives from Section \ref{AnnSec3} in \eqref{eqAnn:DphiJ}, the terms $\mathcal{D}_{\mathbf{E}^*}\mathbf{F}$,
$\mathcal{D}_{\mathbf{E}}\mathbf{F}^*$ and $\mathcal{D}_{\mathbf{F}^*}g$ are null (as $\mathbf{Z}$ and $\mathbf{Z}^*$ are independent variables), so the gradient expressions turn to:
\begin{align}
    \mathcal{D}_{\pmb{\varphi}}J&=(\mathcal{D}_{\mathbf{F}}g)(\mathcal{D}_{\mathbf{E}}\mathbf{F})(\mathcal{D}_{\mathbf{H}}\mathbf{E})\mathbf{H}_{12},\\
    \mathcal{D}_{\pmb{\varphi}^*}J&=(\mathcal{D}_{\mathbf{F}}g)(\mathcal{D}_{\mathbf{E}}\mathbf{F})(\mathcal{D}_{\mathbf{H}^*}\mathbf{E})\mathbf{H}_{12}^*.
\end{align}
If we compute them separately:
\begin{equation*}
    \begin{split}
    &\mathcal{D}_{\pmb{\varphi}}J=\\&=-\text{vec}\left(\mathbf{\Gamma}\right)^T\left(\left(\Ht\Ht^H\right)^{-T}\otimes\left(\Ht\Ht^H\right)^{-1}\right)\left(\Ht^*\otimes\mathbf{I}_N\right)\Ht_{12}\\
    &=-\text{vec}\left(\mathbf{\Gamma}\right)^T\left(\left(\Ht\Ht^H\right)^{-T}\Ht^*\otimes\left(\Ht\Ht^H\right)^{-1}\right)\Ht_{12},
    \end{split}
\end{equation*}
\begin{equation*}
    \begin{split}
        &\mathcal{D}_{\pmb{\varphi}^*}J=\\&=-\text{vec}\left(\mathbf{\Gamma}\right)^T\left(\left(\Ht\Ht^H\right)^{-T}\otimes\left(\Ht\Ht^H\right)^{-1}\right)\left(\mathbf{I}_N\otimes \Ht\right)\mathbf{K}\Ht_{12}^*\\
        &=-\text{vec}\left(\mathbf{\Gamma}\right)^T\left(\left(\Ht\Ht^H\right)^{-T}\otimes\left(\Ht\Ht^H\right)^{-1}\Ht\right)\mathbf{K}\Ht_{12}^*.
    \end{split}
\end{equation*}
Then, using $\text{vec}\left(\mathbf{R}\mathbf{A}\mathbf{S}\right)=\left(\mathbf{S}^T\otimes\mathbf{R}\right)\text{vec}(\mathbf{A})$:
\begin{align*}
    \mathcal{D}_{\pmb{\varphi}}J&=-\text{vec}\left(\left(\Ht\Ht^H\right)^{-T}\mathbf{\Gamma}\left(\Ht\Ht^H\right)^{-T}\Ht^*\right)^T\mathbf{H}_{12},\\
    \mathcal{D}_{\pmb{\varphi}^*}J&=-\text{vec}\left(\Ht^T\left(\Ht\Ht^H\right)^{-T}\mathbf{\Gamma}\left(\Ht\Ht^H\right)^{-T}\right)^T\mathbf{K}\mathbf{H}_{12}^*.
\end{align*}
Since we are to iterate on the phase of the diagonal elements of the RIS, the gradient is finally obtained in \eqref{eqAnn:FinalGradient} (next page).

\begin{figure*}[!t]
\normalsize
\begin{multline}
\label{eqAnn:FinalGradient}
\begin{aligned}
    \mathcal{D}_{\pmb{\phi}}J&=(\mathcal{D}_{\pmb{\varphi}}J)(\mathcal{D}_{\pmb{\phi}}\pmb{\varphi})+(\mathcal{D}_{\pmb{\varphi}^*}J)(\mathcal{D}_{\pmb{\phi}}\pmb{\varphi}^*) \\
    &=-j\text{vec}\left(\left(\Ht\Ht^H\right)^{-T}\mathbf{\Gamma}\left(\Ht\Ht^H\right)^{-T}\Ht^*\right)^T\mathbf{H}_{12}\textup{diag}\left(\pmb{\varphi}\right)+j\text{vec}\left(\left(\Ht\Ht^H\right)^{-H}\mathbf{\Gamma}\left(\Ht\Ht^H\right)^{-H}\Ht\right)^T\mathbf{H}_{12}^*\textup{diag}\left(\pmb{\varphi}^*\right)\\ 
    &=2\operatorname{Im}\left(\text{vec}\left(\left(\Ht\Ht^H\right)^{-T}\mathbf{\Gamma}\left(\Ht\Ht^H\right)^{-T}\Ht^*\right)^T\mathbf{H}_{12}\textup{diag}(\pmb{\varphi})\right)
\end{aligned}
\end{multline}
\hrulefill
\end{figure*}

\section{CHANNEL ESTIMATION IN DIAGONAL RIS-ASSISTED MISO TRANSMISSION}\label{Channel estimation}
\subsection{UPLINK SIGNAL MODEL}
Let us assume that unit power pilot symbols $x_t$ are transmitted by user $k$. The received signal at the $M$ antennas of the BS can be arranged into a column vector:
\begin{equation}
\label{eq:ChannelModelUL}
\begin{aligned}
{\bf{y}}_{tk} &= \sqrt{P_{UL}}\left(\Tilde{\bf{h}}_{bu,k} + \THbr \text{diag}(\pmb{\varphi}_t)\Tilde{\bf{h}}_{ru,k}\right)x_t + {\bf{n}}_{tk}\\
 &= \sqrt {{P_{UL}}} \left( {{{{\bf{\tilde h}}}_{bu,k}} + \THbr \text{diag}\left( {{{\Tilde{\bf{h}}}_{ru,k}}} \right){\pmb{\varphi} _t}} \right){x_t} + {{\bf{n}}_{tk}}\\
 &= \sqrt {{P_{UL}}}\left[\Tilde{\mathbf{h}}_{bu,k} \:\:\:\:\THbr \text{diag}(\Tilde{\mathbf{h}}_{ru,k})\right]\left[ {\begin{array}{c}
1\\{{\pmb{\varphi}_t}}\end{array}} \right]{x_t} + {{\bf{n}}_{tk}}\\
 &= \sqrt {{P_{UL}}} {{\bf{V}}_k}{{\pmb{\varphi}'}_t}{x_t} + 
 {{\bf{n}}_{tk}} \in\mathbb{C}^{M \times 1},
\end{aligned}
\end{equation}
where $P_{UL}$ is the power transmitted in the UL, $\Tilde{\mathbf{h}}_{z,k}$ is the $k$-th column of matrix $\Tilde{\bf{H}}_z=\mathbf{H}_z^T$. We are assuming that the coefficients of the RIS can be selected at every transmitted pilot symbol $j$. If we arrange the signal received along $T$ pilot symbols into a matrix, we obtain:
\begin{equation}
\label{eq: rxSignalUplink}
\begin{aligned}
\mathbf{Y}_k&=\left[{\mathbf{y}_{1k}\:\cdots\:\mathbf{y}_{Tk}}\right]\\&=\sqrt{P_{UL}}\mathbf{V}_k\left[x_1{\pmb{\varphi}'}_1\:\cdots\: x_T{\pmb{\varphi}'}_T\right]+\mathbf{N}_k\\&=\sqrt{P_{UL}}\mathbf{V}_k\mathbf{\Omega}+\mathbf{N}_k\in\mathbb{C}^{M \times T}.
\end{aligned}
\end{equation}
Note that matrix $\mathbf{\Omega}\in \mathbb{C}^{(N_r+1) \times T}$ contains symbols transmitted by the UE and by the RIS simultaneously, and 
\begin{equation}
\mathbf{V}_k=\left[\Tilde{\mathbf{h}}_{bu,k} \:\:\:\: \Tilde{\mathbf{H}}_{br}\text{diag}(\Tilde{\mathbf{h}}_{ru,k})\right] \in \mathbb{C}^{M \times (N_r+1)}
\end{equation}
is the matrix gathering the channel gains from the $k$-th user to the BS. Then, the components in \eqref{eq:channel_model_RIS} can be simply derived from the estimated $\mathbf{V}_k$ as shown in \eqref{eq:H_Vk} (next page).
\begin{figure*}[!t]
\normalsize
\begin{equation}
\label{eq:H_Vk}
\begin{aligned}
\mathbf{H}&=\Hbu+\Hru\PhiM\Hbr=\left(\THbu+\THbr\PhiM\THru\right)^T    \\&=\left(\left[\mathbf{h}_{bu,1}\:\cdots\:\mathbf{h}_{bu,K}\right]+\THbr \text{diag}(\phiM) \left[\mathbf{h}_{ru,1}\:\cdots\:\mathbf{h}_{ru,K}\right]\right)^T  \\&=\left(\left[\mathbf{h}_{bu,1}\:\cdots\:\mathbf{h}_{bu,K}\right]+\THbr\left[\text{diag}(\mathbf{h}_{ru,1})\phiM\:\cdots\:\text{diag}(\mathbf{h}_{ru,K})\phiM\right]\right)^T
\\&=\left(\left[\mathbf{h}_{bu,1}\:\cdots\:\mathbf{h}_{bu,K}\right]+\THbr\left[\text{diag}(\mathbf{h}_{ru,1})\:\cdots\:\text{diag}(\mathbf{h}_{ru,K})\right](\mathbf{I}_M \otimes \phiM)\right)^T
\\&=\begin{bmatrix}
        \mathbf{V}_1^{:,1} & \cdots & \mathbf{V}_K^{:,1}
    \end{bmatrix}^{T}+(\mathbf{I}_M \otimes \phiM)^{T}
    \begin{bmatrix}
        \mathbf{V}_1^{:,2:(N_r+1)} & \cdots & \mathbf{V}_K^{:,2:(N_r+1)}
    \end{bmatrix}^{T}
\end{aligned}
\end{equation}
\hrulefill
\end{figure*}
The notation $\mathbf{V}_k^{:,1}$ stands for the first column of matrix $\mathbf{V}_k$, $\mathbf{V}_k^{:,2:(N_r+1)}$ stands for the submatrix containing columns $2$ through $N_r+1$, and $\pmb{\varphi}$ is the vector of RIS reflection coefficients. Therefore, once $\mathbf{V}_k$ have been estimated, we can obtain $\mathbf{H}$ for some given value of $\phiM$. Another matrix involved in gradient optimization of the RIS reflection coefficients is $\mathbf{H}_{12}$ in \eqref{eq:H12}. It can be obtained from a rearrangement of the elements of $\mathbf{V}_1,...,\mathbf{V}_K$.

\subsection{LEAST SQUARES CHANNEL ESTIMATION}
The least squares estimate of $\mathbf{V}_k$ using the received signal $\mathbf{Y}_k$ is obtained as:
\begin{equation}
\label{least_squares_estimation}
    \hat{\mathbf{V}}_k=\frac{1}{\sqrt{P_{UL}}}\mathbf{Y}_k\mathbf{\Omega}^H\left(\mathbf{\Omega}\mathbf{\Omega}^H\right)^{-1},
\end{equation}
which can be written as $\hat{\mathbf{V}}_k=\mathbf{V}_k+\mathbf{E}_k$, with $\mathbb{E}\{\mathbf{E}_k\} = \mathbf{0}$. By sequentially changing the transmitting user $k = 1,\ldots,K$ it is possible to estimate all involved channels. During the training period, pilot symbols must be transmitted simultaneously from the UE and the RIS. Note that $rank(\mathbf{\Omega})=\min\{T,N_r+1\}$, being $T$ the number of pilot symbols transmitted per user. For $\mathbf{\Omega}\mathbf{\Omega}^H$ to be invertible it is required that $T\geq N_r+1$ \cite{Nadeem20}.

\end{appendices}
\bibliographystyle{ieeetr} 
\bibliography{mybiblio}

\begin{IEEEbiography}[{\includegraphics[width=1in,height=1.25in,clip,keepaspectratio]{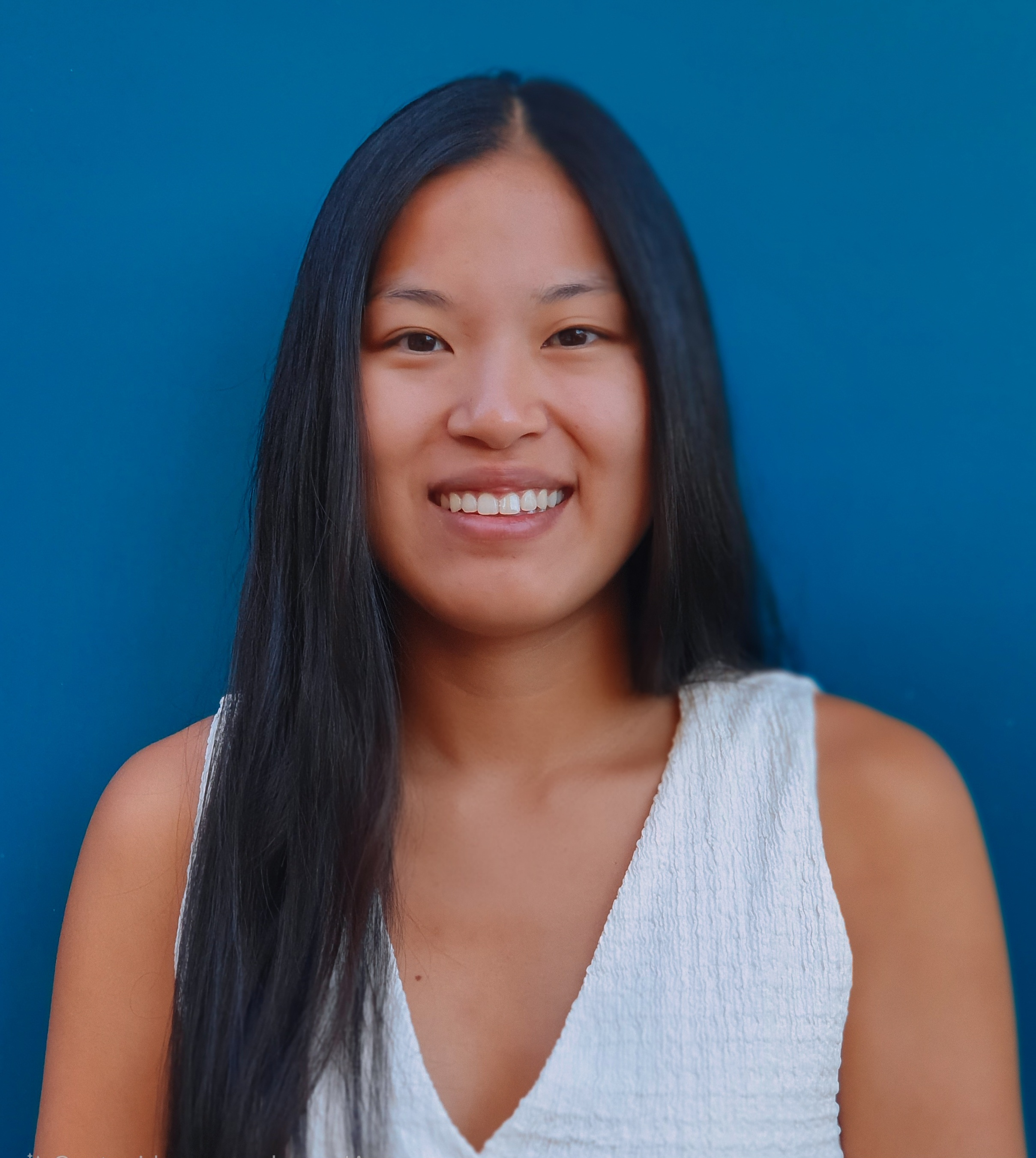}}]{Moreno-Locubiche, Ainna Yue } received the B.S. and M.S degrees in telecommunications engineering from Universitat Politècnica de Catalunya (UPC), Barcelona, Spain, in 2023 and 2025 respectively.
Since 2023 she has been a Research Assistant at the Signal Theory and Communications Department (TSC) of the Universitat Politècnica de Catalunya (UPC). 

Her research interests include advanced wireless communication technologies for 6G networks, with a particular emphasis on the design and optimization of Reconfigurable Intelligent Surfaces (RIS) for signal enhancement to improve coverage, capacity and mobility management in ultra-dense and high-speed network scenarios.
\end{IEEEbiography}

\begin{IEEEbiography}[{\includegraphics[width=1in,height=1.25in,clip,keepaspectratio]{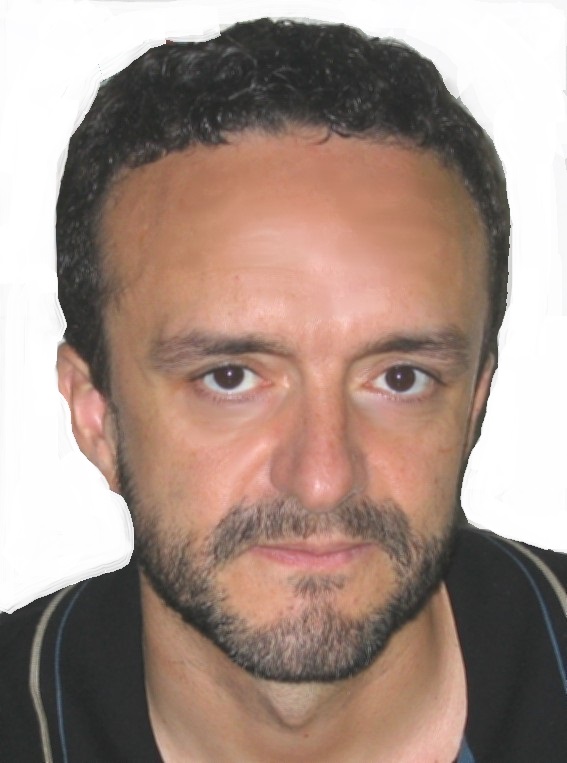}}]{Vidal, Josep } (Senior Member, IEEE) received the M.S. and Ph.D. degree in telecommunication engineering from the Universitat Politècnica de Catalunya (UPC), Barcelona, in 1989 and 1993 respectively. From 1989 to 1990 was with Cognivision Research. Since 2011 he is full professor at the Signal Theory and Communications Department of UPC. 

He has has held research appointments with EPF Lausanne, INP Toulouse, and the University of Hawaii. He is author of 8 book chapters, over 200 articles and 8 inventions on different aspects of statistical signal and array processing, information and communication theory and machine learning. 

Since 2002, he has coordinated collaborative EC-funded projects ROMANTIK, ROCKET, FREEDOM, TROPIC, TUCAN3G, 5GSmartFact, all in different areas of MIMO and relay communications, self-organization, cooperative transmission, heterogeneous networks and MEC. He has co-organized 4 international workshops and has served as an Associate Editor of the IEEE Transactions on Signal Processing.
\end{IEEEbiography}

\begin{IEEEbiography}[{\includegraphics[width=1in,height=1.25in,clip,keepaspectratio]{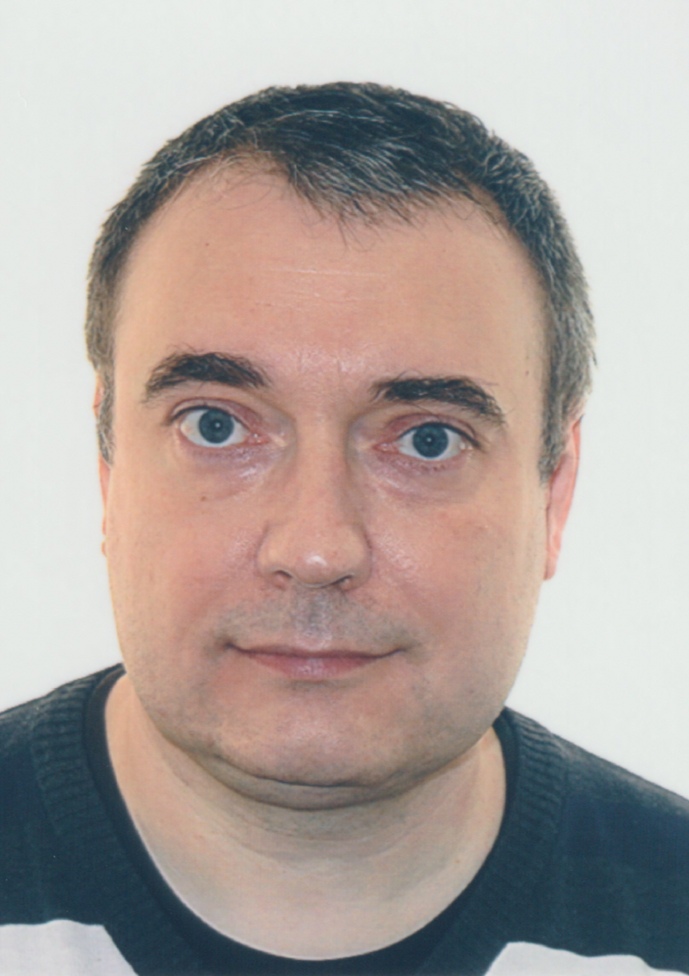}}]{Pascual-Iserte, Antonio } (Senior Member, IEEE) was born in Barcelona, Spain, in 1977. He received the degree in electrical engineering and the Ph.D. degree from the Universitat Politècnica de Catalunya (UPC), Barcelona, in 2000 and 2005, respectively. From 1998 to 1999, he was a Teaching Assistant with the Department of Electronic Engineering, UPC. From 1999 to 2000, he was with Retevision R\&D, working on the implantation of the DVB-T and T-DAB networks in Spain. In 2001, he joined the Department of Signal Theory and Communications, UPC, where he worked as a Research Assistant until 2003. He became an Assistant Professor in 2003, Associate Professor in 2008, and Full Professor in 2024. He was also regular collaborator with the Centre Tecnològic de Telecomunicacions de Catalunya (CTTC) from 2005 to 2012. 

He has been involved in several research projects funded by the Spanish Government and the European Commission. He has also published several papers in international and national conference proceedings and journals. His current research interests include array processing, robust designs, orthogonal frequency-division multiplexing, MIMO channels, multiuser access, stochastic geometry, 5G network-scale performance evaluation, massive machine-type communications, mmWave channel modeling, delay-Doppler communications, and optimization theory.

Dr. Pascual-Iserte received the First National Prize of 2000/2001 University Education from the Spanish Ministry of Education and Culture and the Best 2004/2005 Ph.D. Thesis Prize from UPC.  
\end{IEEEbiography}

\begin{IEEEbiography}[{\includegraphics[width=1in,height=1.25in,clip,keepaspectratio]{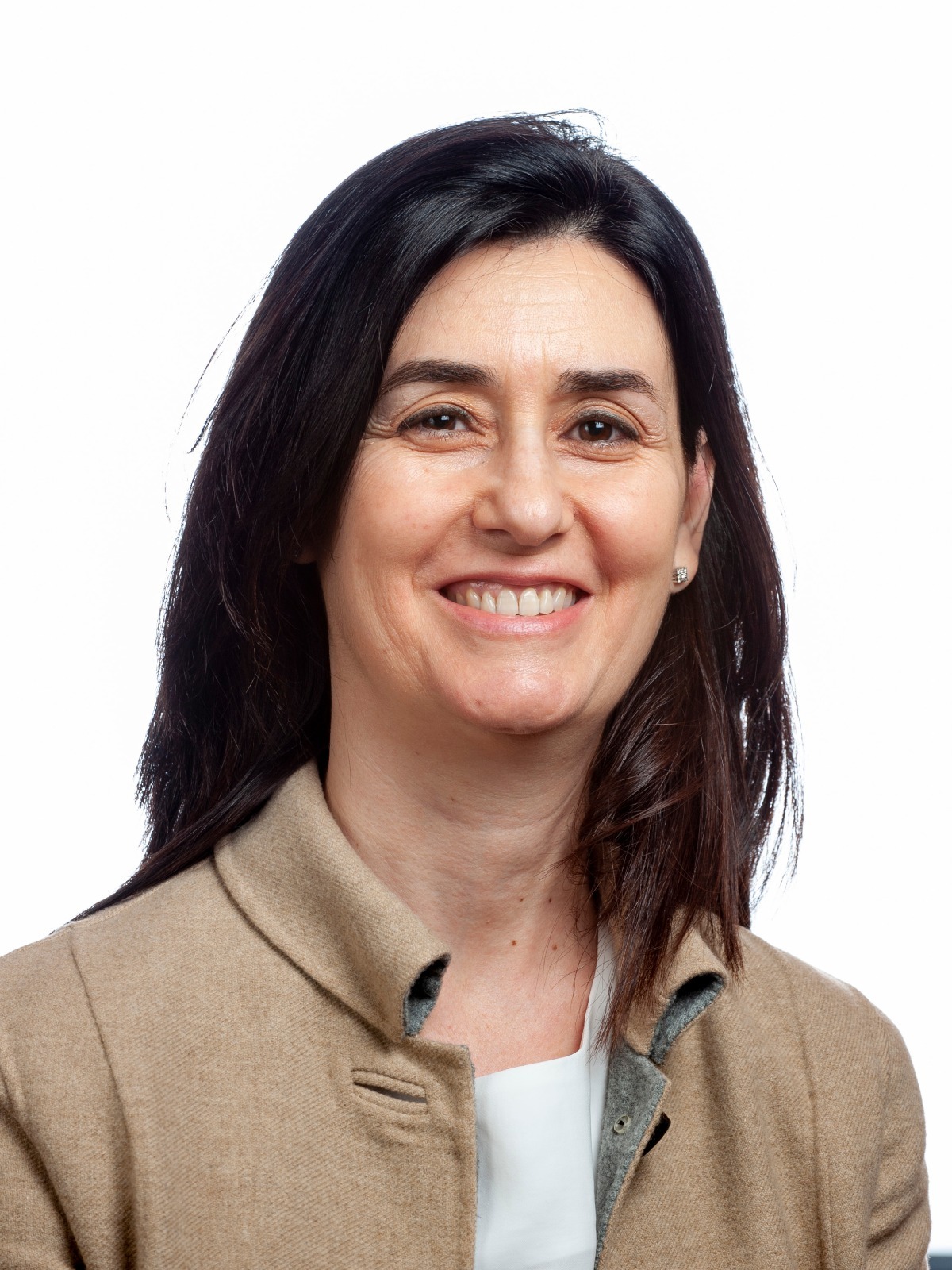}}]{Muñoz, Olga } (Senior Member, IEEE) received M.S. and Ph.D. degrees in electrical engineering from the Universitat Politècnica de Catalunya (UPC), Spain, in 1993 and 1998, respectively. In 1994, she joined the Department of Signal Theory and Communications, UPC, where she teaches graduate and undergraduate signal processing and communications courses. 

She accumulates substantial experience in relaying and cooperative upgraded networks backed by her work on
European Commission projects ROMANTIK (5thFP), FIREWORKS (6thFP), ROCKET (7thFP), TROPIC (7thFP) and in TUCAN3G (7thFP). Also in the Spanish Government funded projects MOSAIC (call 2010) and 6-SENSES (call 2022). 
More recently, she has been designing, analyzing, and evaluating radio technologies in ultra-dense networks to meet the capacity and quality of service requirements, distributed intelligence, and flexibility needed for 6G systems. She has published over 80 papers in books, international conferences, and journals in the areas of signal processing and communications.

She has served as a reviewer for the Spanish Research Council and in numerous journals and conferences. 
\end{IEEEbiography}

\end{document}